\documentclass[twocolumn,preprintnumbers,amsmath,amssymb]{revtex4}
\usepackage{graphicx}
\usepackage{gensymb}
\usepackage{bm}
\usepackage{amsmath}
\usepackage{amssymb}
\usepackage{amsthm}
\usepackage{amsfonts}
\usepackage{enumerate}
\usepackage{comment}
\usepackage{centernot}
\usepackage{tikz}
\usepackage{braket}
\usepackage{subfigure}

\begin{document}

\title{Moir{\'e} semiconductors on the twisted bilayer dice lattice}
 
 \author{Di Ma$^1$}
 \thanks{These two authors contribute equally.}
 \author{Yu-Ge Chen$^{2\dagger}$}
 \thanks{These two authors contribute equally.}
 \author{ Yue Yu$^{1}$}
 \author{ Xi Luo$^{3}$}
 \thanks{Correspondence to: chenyuge@iphy.ac.cn, xiluo@usst.edu.cn}
 \affiliation {  1. Department of Physics, Fudan University, Shanghai 200433,China\\ 
 	2. Beijing National Laboratory for Condensed Matter Physics and Institute of Physics, Chinese Academy of Sciences, Beijing 100190, China\\
 	3. College of Science, University of Shanghai for Science and Technology, Shanghai 200093, China\\}

 \date{\today}
 
 \begin{abstract}
 	We propose an effective lattice model for the {moir{\'e}} structure of the twisted bilayer dice lattice. {In the chiral limit, we find that there are flat bands at the zero-energy level at any twist angle besides the magic ones, and these flat bands are broadened by small perturbation away from the chiral limit.} The flat bands  contain both bands with zero Chern number which originate from the destructive interference of {the states on} the dice lattice and the topological nontrivial bands at the magic angle.
	  The existence of the flat bands can be detected from the peak-splitting structure of the optical conductance at all angles, while the transition peaks do not split and only occur at magic angles in twisted bilayer graphene.
 \end{abstract}
 
\maketitle

\section{ Introduction.}

The search for a flat-band system has become one of the new trends over the past few decades. Due to the large effective mass of the quasi particles {in the flat-band system}, the density of states (DOS) is high,
and the kinetic energy of the carriers is strongly quenched. Therefore, the flat-band system is a good candidate for studying strongly correlated electronic states induced by strong Coulomb interaction, such as ferromagnetism \cite{JOP1991,PRA2010,PRL1992}, heavy fermions \cite{nature2020,np2019}, fractional Chern insulators \cite{IJMP2013,PRL2011}, Wigner crystals \cite{PRL2007,PRB2008}, and unconventional superconductivity \cite{RMP1990,PCS2007,IOS2020}. 

Traditionally, the nearly-flat-band system can be achieved by invoking fine-tuned nearest-neighbor hoppings or long-ranged hoppings or by breaking time-reversal symmetry. 
Several lattice models have been proposed along these lines in kagome \cite{PTP1951}, Lieb \cite{PRL1989}, and dice lattices \cite{PRB1986}. The existence of the flat band is guaranteed by the destructive interference of the Wannier functions of the lattice structure, and the flat-band states are identified as compact localized states \cite{PRL2007,PRB2008,PRB1986,PRL2013,CPB2014,sathe1,sathe2}. This destructive interference protection can also be generalized to lattices with mirror symmetry \cite{chen2022}. 
Usually, {this kind of} flat band has a zero Chern number in the lattice model with nearest neighbor hopping only \cite{sathe3}. On the dice lattice, the flat band can acquire a non-zero Chern number by invoking Rashba spin-orbital coupling and exhibits an anomalous quantum Hall effect by adding onsite Hubbard interactions, which could be realized in the transition-metal oxide SrTiO3/SrIrO3/SrTiO3 trilayer heterostructure by growing in the (111) direction \cite{wang2011}. 
Materials with flat-band structures along these lines have also been reported in Cu(111) confined by CO molecules \cite{np2017}, optical lattices, and cold-atom systems \cite{shen2010,njp2014,prl2015a,prl2015b,sa2015,ol2016}. The topology of the {dice} lattice with non-Hermiticity has also been studied \cite{sarkar2023}. In three dimensions, a famous example is the Kane semimetal, where the flat band structure is associated with the triplet degenerate nodes and can be described by a three-dimensional Lieb lattice model \cite{Luo2018}. The low-energy quasi particle, the Kane fermion, can be viewed as a fermionic photon, i.e., a spin-1 fermion. {The effective Hamiltonian near the triplet degenerate node is $H_{jk}=\epsilon_{ijk}p_i$, where $\epsilon_{ijk}$ is the totally anti-symmetric tensor, and $p_i$ is the momentum. By squaring the Hamiltonian, $H^2\sim p_ip_j-p^2\delta_{ij}$, which is related to the Hamiltonian of the photon. The case here is similar to squaring the Dirac Hamiltonian to obtain the Hamiltonian for the Klein-Gordon equation. Therefore, the Kane fermion can be viewed as a  fermionic photon} and the flat band of the Kane fermion corresponds to the longitudinal mode of the photon. Experimentally, the {spinful} Kane fermion has been reported in  Hg$_{1-x}$
	Cd$_x$Te \cite{hgcdte} and Cd$_3$As$_2$ \cite{cdas}. The existence of the flat band is shown in the optical conductance by the large peaks near zero frequency \cite{hgcdte,Luo2019}. {The spinless Kane fermion is proposed to exist in materials with space groups 199 and 214, such as Ag$_3$Se$_2$Au and Pd$_3$Bi$_2$S$_2$ \cite{science2016}. Band structures with triple nodal points have also been proposed in ZrTe \cite{PRX2016}, LaPtBi \cite{PRL2017}, and  $A$Pd$_3$($A$=Pb, Sn) \cite{PRB2018}. More kinds of topological materials would be found by the method of symmetry indicators and topological quantum chemistry \cite{n1,n2,n3}.}



A new mechanism for generating a flat band was found in twisted bilayer graphene (TBG) at a magic twist angle $\theta\sim 1.08^{\degree}$ \cite{Bistritzer-MacDonald2011,mac2020}. 
Unlike the destructive-interference-induced flat band, the flat-band structure of TBG originates {from} the extremely large band folding of the {moir{\'e}} structure, and TBG becomes a strongly correlated electron system. 
Very soon thereafter, superconductivity was reported in TBG \cite{cao2018,bernevig2018}. The flat band in TBG has a non trivial Chern number \cite{bernevig2020}, which can be explained by the zeroth chiral Landau levels of Dirac/Weyl fermions \cite{Liu2019Pseudo}.
 Away from the half filling, the fractional Chern insulator phase was also proposed and reported in TBG \cite{Tarnopolsky2019,ashvin2020,ashvin2021} and the twisted bilayer MoTe$_2$ \cite{xu2023,xu2023b,shan2023}.



In this paper we consider the combination of {moir{\'e}} structure and destructive-interference-induced flat bands, namely, a twisted bilayer dice lattice (TBD). {Inspired by Bistritzer and MacDonald's continuum lattice model for TBG \cite{Bistritzer-MacDonald2011},} we construct a lattice model for the TBD in the reciprocal space. {We find that there are flat bands with zero energy in the chiral limit at any twist angle besides the magic ones that are broadened by small perturbation away from the chiral limit.} The flat bands are contributed from the ones with zero Chern number as in the Dice lattice and the non trivial ones at the magic angle; therefore, the TBD is a playground for studying the interplay between the zero-Chern-number flat bands and non trivial ones. We further confirm this scenario by considering the pseudo-Landau-level description \cite{Liu2019Pseudo} {and its optical conductance.}

The paper is organized as follows. {In Sec. \ref{secii}, we provide the detailed construction of the lattice model of the TBD. We also {introduce} the concept of chiral limit to the TBD. From the chiral limit of the TBD, we show the origin of the flat bands in the TBD.} {There are flat bands in the TBD in the chiral limit at all angles other than the magic ones.} We also numerically calculate the Bloch band structure, and compare it with that of TBG. In Sec. \ref{seciv} we use the pseudo-Landau-level language to describe the physics of the flat bands in the TBD, where the pseudo-magnetic field is caused by the interlayer hopping. We can directly find that, besides the topological zeroth Landau level, the higher Landau levels which are {topologically} trivial also contribute to the flat bands. In Sec. \ref{secv} we use the degenerate perturbation method to calculate the optical conductance of the TBD and we find that the flat bands contribute a peak-splitting structure, which is conclusive evidence of the experimental prediction for the existence of the flat bands. This phenomenon also exists at all angles besides the magic ones when compared with TBG. Section \ref{con} provides a brief summary and discussion of our conclusions.


\section{ Effective model of the TBD} \label{secii}

\begin{figure}
	\centering
	\subfigure[\label{a}]{\includegraphics[scale = 0.3]{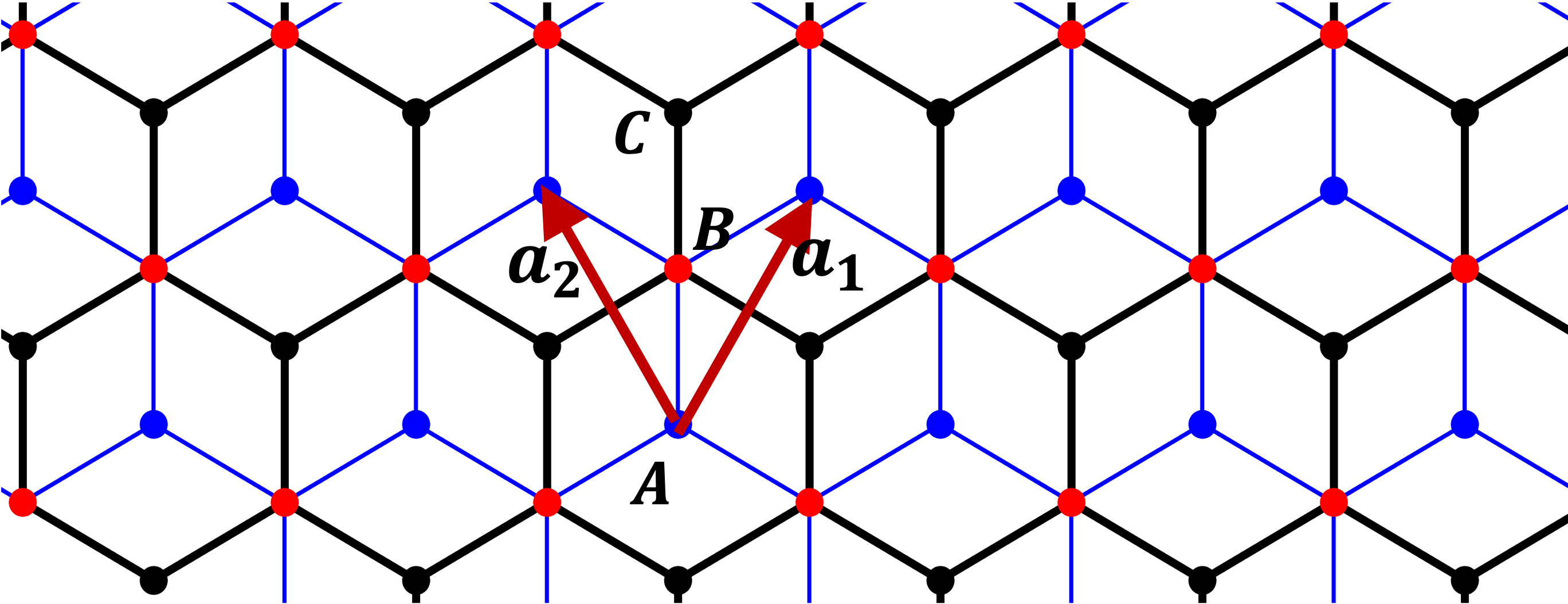}}\\
	\subfigure[\label{b}]{\includegraphics[scale = 0.4]{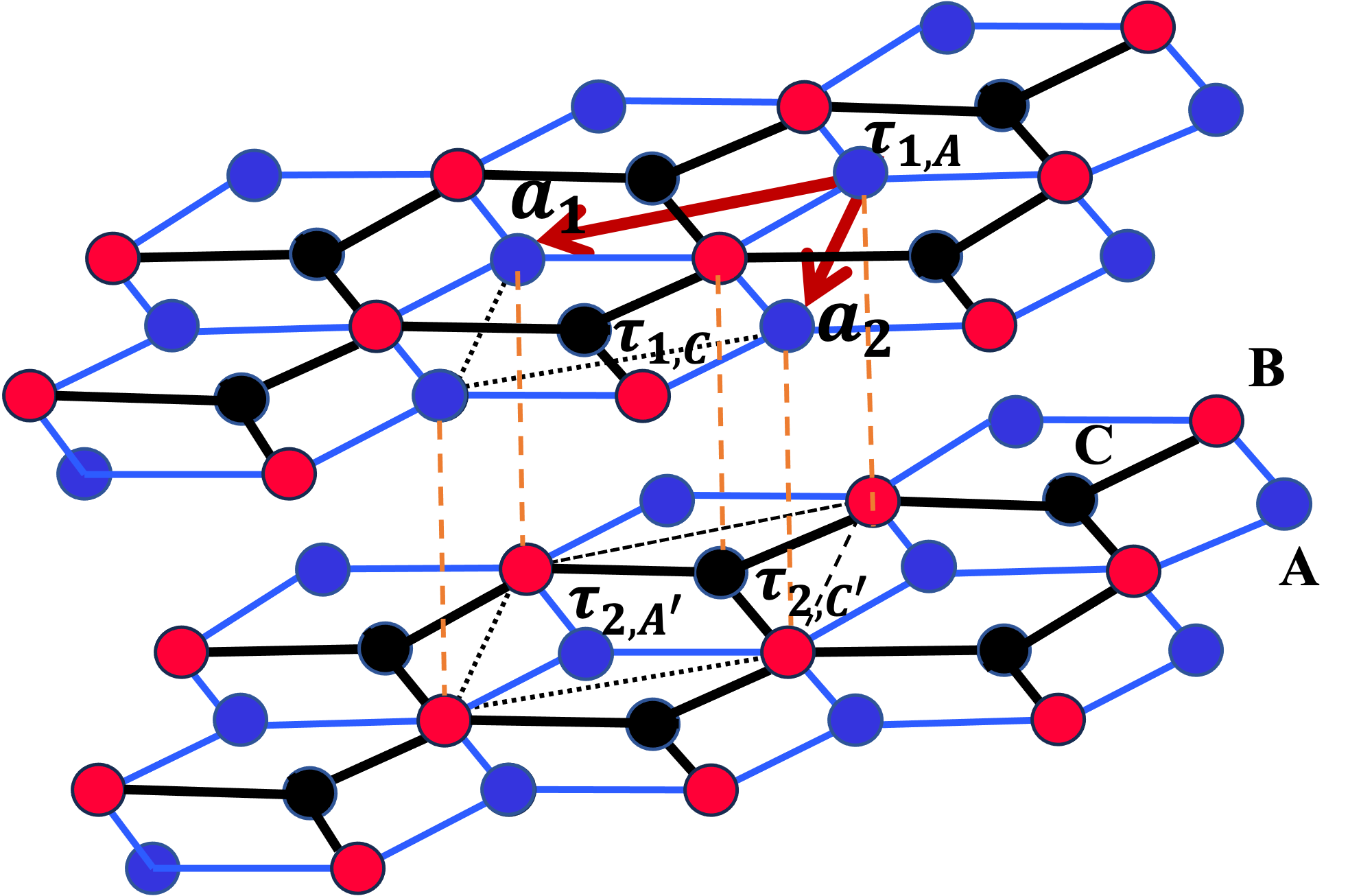}}
	\caption{(color online)
		{ (a) Structure of the dice lattice. Blue, red, and black dots correspond to $A$, $B$, and $C$ sites, respectively. The lattice vectors are denoted by $\boldsymbol{a}_2$ and $\boldsymbol{a}_2$.  
		 (b) Bilayer dice lattice in $A-B$ (Bernal) stacking. The vector ${\bf \tau}$ labels the position of the sublattice site in a unit cell. 
		}
		\label{fig1}	}
\end{figure}

\begin{figure}
	\centering
	\subfigure[\label{a}]{\includegraphics[scale = 0.5]{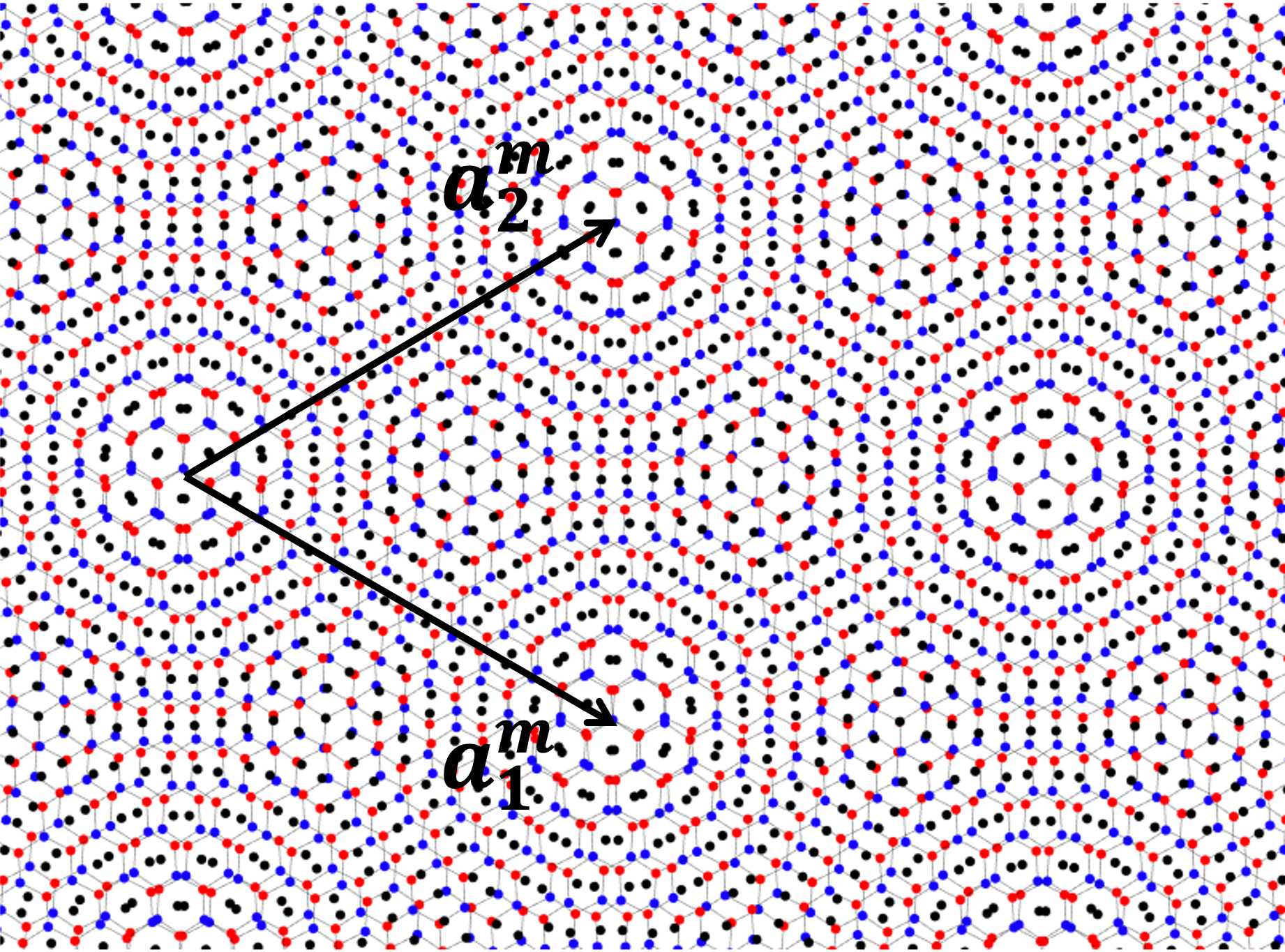}}
	\subfigure[\label{b}]{\includegraphics[scale = 0.4]{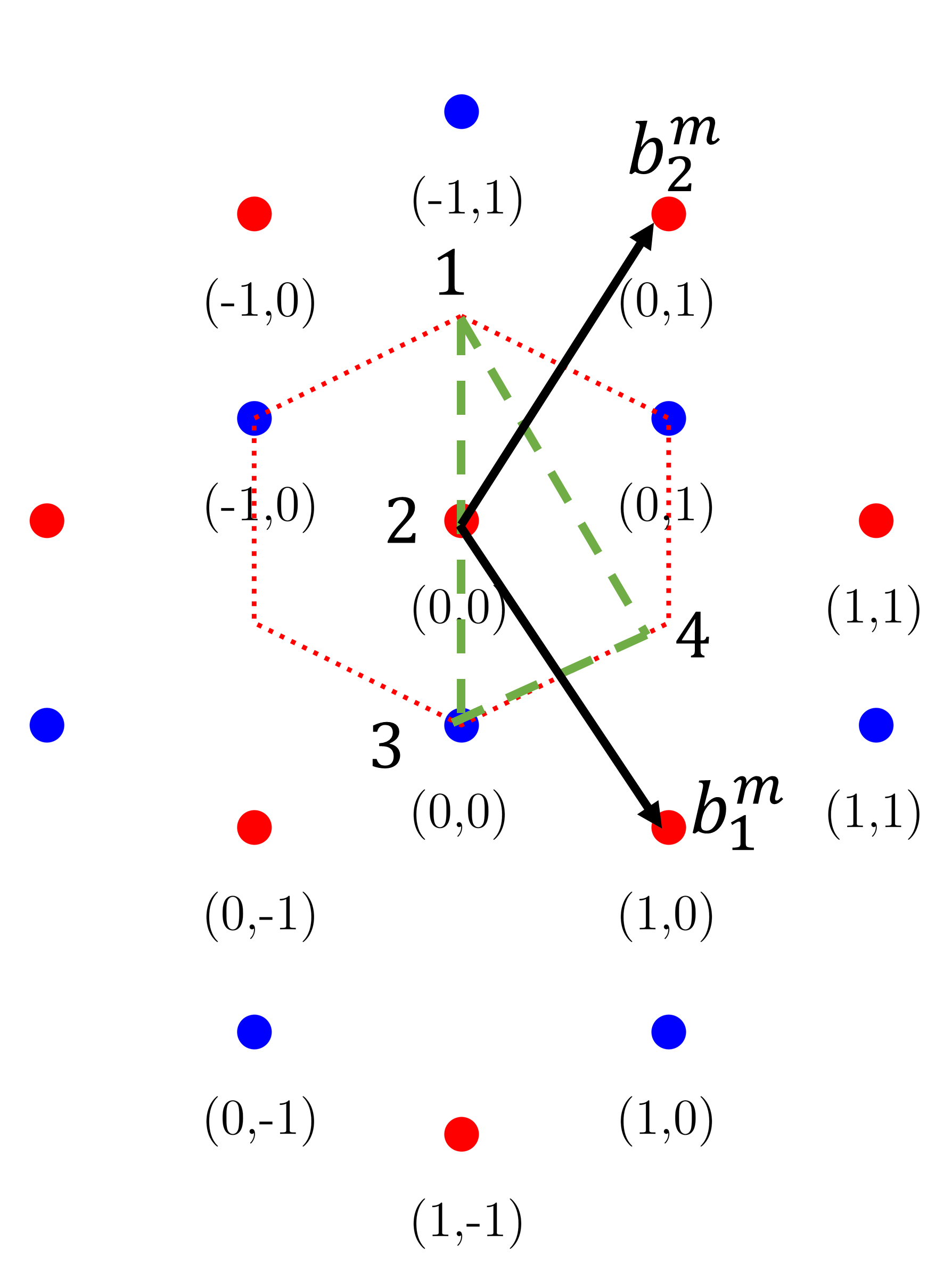}}
	\caption{(color online)
		{ 
			(a) {Moir{\'e}} period structure of the TBD with unit lattice vectors $\boldsymbol{a}_1^m = L_s(\frac{\sqrt{3}}{2}, -\frac{1}{2} )$ and $\boldsymbol{a}_2^m = L_s(\frac{\sqrt{3}}{2}, \frac{1}{2} )$,
			where $L_s = \frac{a}{2\sin (\theta/2)}$ is the size of the {moir{\'e}} unit cell.
			(b) Reciprocal space of the TBD with reciprocal vectors $\boldsymbol{b}^m_1 = \frac{4\pi}{\sqrt{3}L_s}(\frac{1}{2}, -\frac{\sqrt{3}}{2})$ and $\boldsymbol{b}^m_2 = \frac{4\pi}{\sqrt{3}L_s}(\frac{1}{2}, \frac{\sqrt{3}}{2})$. The coordinate $(l_1,l_2)$ means the vector ${\bf b}=l_1\boldsymbol{b}^m_1+l_2\boldsymbol{b}^m_2$. The red dashed line marks the first Brillouin zone with green dashed lines marking the reciprocal path for calculating the TBD band structure. 
		}
		\label{fig3}	}
\end{figure}

	The dice lattice can be viewed as two honeycomb lattices ($A-B$ and $B-C$) sharing the same sublattice site $B$. {The lattice base vectors are $\boldsymbol{a}_1 = a(\frac{1}{2}, \frac{\sqrt{3}}{2} )$ and $\boldsymbol{a}_2 = a(-\frac{1}{2}, \frac{\sqrt{3}}{2} )$ and the corresponding reciprocal vectors
		$\boldsymbol{b}_1 = \frac{4\pi}{\sqrt{3}a}(\frac{\sqrt{3}}{2}$, and $\frac{1}{2} ), \boldsymbol{b}_2 = \frac{4\pi}{\sqrt{3}a}(-\frac{\sqrt{3}}{2}, \frac{1}{2} )$, where $a = \sqrt{3}d$ is lattice constant, and $d$ is the distance between two nearest sites [see Fig. \ref{fig1}(a)].} For simplicity, we consider a spinless lattice model with nearest-neighbor hopping only. Due to this similarity, the TBD has the same {moir{\'e}} structure as TBG {[see Fig. \ref{fig3}(a)]}. There is a flat band $E=0$ in the lattice spectrum. The low energy behavior near the Dirac point can be captured by $H_{eff}={\bf k \cdot S}$, where ${\bf k}=(k_1,k_2,0)$ is the lattice momentum, and ${\bf S}$ is the spin-1 generalization of the Pauli matrix that acts on sublattice space {of the order of $A$, $B$, and $C$},
\begin{gather}
	S_1 = \begin{pmatrix}
		0& 1& 0\\
		1& 0& 1\\
		0& 1& 0
	\end{pmatrix},\quad
	S_2 = 
	\begin{pmatrix}
		0& -i& 0\\
	i&0& -i\\
		0& i &0
	\end{pmatrix},\quad
	S_3 = 
	\begin{pmatrix}
	1& 0& 0\\
		0& 0& 0\\
		0& 0 & -1
	\end{pmatrix}. \label{s-matrix}
\end{gather} 

Since the dice lattice can be realized in the SrTiO3/SrIrO3/SrTiO3 trilayer heterostructure by growing in the (111) direction \cite{wang2011}, it is possible to realize the TBD in this material by twisting when growing. 

For TBG, Bistritzer and MacDonald proposed a low-energy effective continuum Dirac model of the {moir\'e} structure for a small twist angle $\theta<10^{\degree}$ {using Bloch bands near the Dirac points}. The effective model consists of two isolated graphene layers and hopping terms between them. With this model they reveal flat Bloch bands in the electric structure at magic twist angles which give rise to a high DOS \cite{Bistritzer-MacDonald2011}.
Similar to the case of TBG, we follow Ref. \cite{Bistritzer-MacDonald2011} to construct a continuum model for the TBD. {In principle, for the Bloch-band-based effective theory to be valid, the valley structure should be present. However, the global flat band in the single-layer dice model may negate this validity. This obstacle {could} be avoided by including the second-nearest-neighbor hopping in the lattice model such that the global flat bands become dispersive and acquire the valley structure. In a recent paper \cite{arv2023} Zhou $et$ $al.$ confirmed that the flat-band structure is substantiated in the TBD in the absence of the second-nearest-neighbor hopping. Therefore, we can only consider the nearest-neighbor hopping in the TBD for simplicity. {Later we will show that this is the case in the chiral limit of the continuum model, and there are exact flat bands at all angles. In contrast, away from the chiral limit, the results would be less predictive if the second-nearest-neighbor hopping were zero and the valley structure were absent due to the flat bands.}}

In the TBD, we keep the top layer 1 fixed and rotate the bottom layer 2 by $\theta$ with respect to layer 1. The effective TBD Hamiltonian contains intralayer and interlayer parts. The low-energy intra-Hamiltonian reads
\begin{gather}
	H = v_f\sum_k \psi_{1,\mathbf{k}}^\dagger \mathbf{S \cdot k } \psi_{1, \mathbf{k}} + 
	v_f\sum_k \psi_{2,\mathbf{k}}^\dagger \mathbf{S_{\theta} \cdot k } \psi_{2, \mathbf{k}}, \label{eq2}
\end{gather}
where {$\psi_{1,2}$ are the annihilation operators in layers 1 and 2 respectively,} and ${\bf S}_\theta = e^{(i\theta/2) S_3} (S_1, S_2)e^{(-i\theta/2) S_3}$.

For the interlayer {hopping} term $H_{\bot}$, we consider nearest-neighbor hopping from layer 1 in the $\alpha$ $(\alpha = A, B, C)$ 
 sublattice to the closest sublattice $\beta$ $(\beta = A', B', C')$ in layer 2 (see Fig. \ref{fig1}). {The hopping amplitude depends on the difference between the two sites. We have,
 	\begin{gather}\label{aa}
 		H_{\bot} = \sum_{\mathbf{k}_1, \mathbf{k}_2, \alpha, \beta} \psi_{1, \alpha}^\dagger(\mathbf{k}_1) T^{\alpha, \beta}_{1,2}(\mathbf{k}_1, \mathbf{k}_2)\psi_{2, \beta}(\mathbf{k}_2)+ H.c,
 	\end{gather}
 with
 \begin{eqnarray}
 	&&T^{\alpha, \beta}_{1, 2}(\mathbf{k}_1, \mathbf{k}_2) \nonumber\\
 	&=& \frac{1}{A_{u.c}}\sum_{\mathbf{G}_1, \mathbf{G}_2}
 	e^{-i\mathbf{G}_1 \times \tau_{1, \alpha}}t({\bf K}_1 + \mathbf{k}_1 + \mathbf{G}_1) \cdot \nonumber\\
 	&&e^{-i\mathbf{G}_2 \cdot \tau_{2, \beta}}
 	\delta_{{\bf K}_1 + \mathbf{k}_1 + \mathbf{G}_1, {\bf K}_2 + \mathbf{k}_2 + \mathbf{G}_2}, \label{hop}
 \end{eqnarray}
where $A_{u.c}$ is unit cell area and $K_{1,2}$ represent the Dirac point for each layer which satisfies $K_2 = R_\theta \cdot K_1$, where $R_\theta$ is the rotation matrix. Here $\tau_{1/2, \alpha / \beta}$ is a vector connecting the two sites in the unit cell. For the TBD, we consider the $A-B$ stacking (Bernal) configuration coordinates as 
$\tau_{1,A} = \tau_{2, B'} = (0, 0)$, $\tau_{1,B} = \tau_{2, C'} = (0, d)$, and $\tau_{1, C} = \tau_{2, A'} = (0 ,2d)$ [see Fig. \ref{fig1}(b)]. 
To compare the results with TBG, we choose $ d = 1.42$ $\text{\AA}$, the lattice constant of graphene. Here $t({\bf k})$ is the Fourier transformation of the tunneling amplitude {$t({\bf r})$} which satisfies  $t^{\alpha, \beta}_{1,2}(\mathbf{r}_1, \mathbf{r}_2) = t^{\alpha, \beta}_{1,2}(\mathbf{r}_1 + \tau_{1, \alpha} -\mathbf{r}_2 - \tau_{2, \beta})$ and decays rapidly if $k$ in reciprocal space exceeds the Dirac point \cite{Bistritzer-MacDonald2011}. Considering  this 
property we only need to choose three vectors $\mathbf{G}_l = \mathbf{g}_{(l), 1}, \mathbf{g}_{(l), 2},\mathbf{g}_{(l), 3}$, with $\mathbf{g}_{(l), 1} = 0$, $\mathbf{g}_{(l), 2} = \mathbf{b}_{(l), 2}$, and $\mathbf{g}_{(l), 3} = -\mathbf{b}_{(l), 1}$ the reciprocal lattice vectors, $(l) = (1),(2)$ the layer index, and $ \mathbf{b}_{(1), i}=R_\theta \mathbf{b}_{(2), i}$. Substituting these three vectors into the hopping matrix (\ref{hop}), we have
\begin{eqnarray}
&&	T_{1,2}({\bf k}_1,{\bf k}_2)\nonumber\\
&=&T_{\bm{q}_b}\delta_{{\bf k}_1-{\bf k}_2-\bm{q}_b}+T_{\bm{q}_{tr}}\delta_{{\bf k}_1-{\bf k}_2-\bm{q}_{tr}}+T_{\bm{q}_{tl}}\delta_{{\bf k}_1-{\bf k}_2-\bm{q}_{tl}},\nonumber\\
\end{eqnarray} 
where $\bm{q}_b = \frac{8\pi \sin(\theta/2)}{(3a)}(0, -1)$, $\bm{q}_{tr} = \frac{8\pi \sin(\theta/2)}{(3a)}(\frac{\sqrt{3}}{2}$, and $\frac{1}{2}), \bm{q}_{tl} = \frac{8\pi \sin(\theta/2)}{(3a)}(-\frac{\sqrt{3}}{2}, \frac{1}{2})$ are the vectors connecting the nearest Dirac points of the two layers in the moir{\'e} Brillouin zone [see Fig. \ref{fig3}(b)],} 
and
\begin{eqnarray}
&&	T_{\bm{q}_b} = W\begin{pmatrix}
		1& 1& 1\\
		1& 1& 1\\
		1& 1& 1
	\end{pmatrix},\nonumber\\
&&	T_{\bm{q}_{tr}} = W e^{-i\mathbf{g}_{1,2}\cdot \mathbf{\tau}_0}
	\begin{pmatrix}
		e^{i\phi}& 1& e^{-i\phi}\\
		e^{-i\phi}& e^{i\phi}& 1\\
		1& e^{-i\phi} & e^{i\phi}
	\end{pmatrix},\nonumber\\
&&	T_{\bm{q}_{tl}} = W e^{-i\mathbf{g}_{1,3}\cdot \mathbf{\tau}_0}
	\begin{pmatrix}
		e^{-i\phi}& 1& e^{i\phi}\\
		e^{i\phi}& e^{-i\phi}& 1\\
		1& e^{i\phi} & e^{-i\phi}
	\end{pmatrix}, \label{transition}
\end{eqnarray} 
{where $W = \frac{t(|K|)}{A_{u.c}}$ with $|K| = 4\pi/(3\sqrt{3}d)$, and $\phi = 2\pi/3$. Here we choose $W= 110$ meV as in graphene \cite{Bistritzer-MacDonald2011}. We choose also $\tau_0 = 0$, which is the translation vector of the TBD. For later convenience, we also denote the transition amplitude between the two layers by {$W_{\alpha\beta}$}. } 

\subsection{Chiral limit of the TBD}

{The origin and topological nature of the flat bands can be revealed in the chiral limit. In the Ref. \cite{Tarnopolsky2019}, {Tarnopolsky $et$ $al$}. proposed the chirally symmetric continuum model for TBG, which is also known as the chiral limit. In their model they considered a Hamiltonian
	\begin{equation*}
		H_{TBG}=\left(
		\begin{array}{cc}
			0	& \mathcal{D}^*(-{\bf r})  \\
			\mathcal{D}({\bf r})	& 0
		\end{array}
		\right), \quad
		\mathcal{D}({\bf r})=\left(
		\begin{array}{cc}
			-i\bar{\partial}	&  \alpha U({\bf r})\\
			\alpha U({\bf r})	& -i\bar{\partial}
		\end{array}
		\right),
	\end{equation*}
	with the basis $\Phi({\bf r})=(\psi_1,\psi_2,\chi_1,\chi_2)^T$, where $1$ and $2$ are the layer indices and $\psi$ and $\chi$ correspond to the sublattice. Here $\alpha$ is a parameter, $U({\bf r})$ is the interlayer potential, and $\bar{\partial}=\partial_x+i\partial_y$. The chiral symmetry is manifested by the particle-hole symmetry $\{H,\sigma_z\otimes 1 \}=0$, where $\sigma_z$ acts in the sublattice space. The flat bands in TBG satisfy $\mathcal{D} \psi_{\bf k}({\bf r})=0$, which also determines the magic angle, and the flat-band wave function $\psi_{\bf k}({\bf r})$ behaves like the one in the quantum Hall effect on a torus \cite{Tarnopolsky2019}. Therefore, these flat bands in TBG are topological. Another reason is that the flat bands can be explained as the zeroth Landau level of the Weyl fermion under a pseudomagnetic field which comes from the lattice distortion of the twisting \cite{Liu2019Pseudo} and the zeroth Landau level of the Weyl fermion is topological.}


{Inspired by this model, we generalize the chiral limit to the TBD model by substituting from the Pauli matrices that act on the sublattice space to the $3\times 3$ ${\bf S}$ matrices defined in (\ref{s-matrix}),
	\begin{equation}
		H_{TBD}=\left(
		\begin{array}{ccc}
			0	& \mathcal{D}^*(-{\bf r}) & 0  \\
			\mathcal{D}({\bf r})	& 0 & \mathcal{D}^*(-{\bf r})\\
			0& 	\mathcal{D}({\bf r}) & 0
		\end{array}
		\right),\label{htbd}
	\end{equation}
	with the particle-hole symmetry {$\{H_{TBD},S_3\otimes 1 \}=0$}. The basis is now $\Psi=(\psi^A_1,\psi^A_2,\psi^B_1,\psi^B_2,\psi^C_1,\psi^C_2)^T$, where $\psi^{A,B,C}$ labels the sublattice. Therefore, the generalized chiral symmetry in the TBD corresponds to choosing $W_{AA'} = W_{BB'} = W_{CC'} = W_{AC'} = W_{CA'}= 0$, and the hopping matrices defined in (\ref{transition}) become
\begin{eqnarray}
	&&	T_{\bm{q}_b}^c = W\begin{pmatrix}
			0& 1& 0\\
			1& 0& 1\\
			0& 1& 0
		\end{pmatrix},\nonumber\\
	&&	T_{\bm{q}_{tr}}^c = W e^{-i\mathbf{g}_{1,2}\cdot \mathbf{\tau}_0}
		\begin{pmatrix}
			0& 1& 0\\
			e^{-i\phi}& 0& 1\\
			0& e^{-i\phi} & 0
		\end{pmatrix},\nonumber\\
	&&	T_{\bm{q}_{tl}}^c = W e^{-i\mathbf{g}_{1,3}\cdot \mathbf{\tau}_0}
		\begin{pmatrix}
			0& 1& 0\\
			e^{i\phi}& 0& 1\\
			0& e^{i\phi} & 0
		\end{pmatrix}. 
\end{eqnarray}}

{Now we can count the number of flat bands. The usual band counting from TBG at the magic angle is one flat band per valley per spin, with a total number of four,  {from each of} $\mathcal{D}\psi=0$ and $\mathcal{D}^*\psi=0$. In TBD, there are two kinds of flat bands. One is similar to the case of TBG, which comes from $\mathcal{D}\psi_A=0$ and $\mathcal{D}^*\psi_C=0$. They produce four flat bands  {each}. The $A$ and $C$ sublattice indices correspond to the valley indices in TBG. For $\psi_B$, $\mathcal{D}\psi_B=0$ and $\mathcal{D}^*\psi_B=0$ should be satisfied for flat bands. In the $\alpha=0$ limit, this means $\psi_B$ should be both holomorphic and antiholomorphic, which is a constant if $\psi_B$ has no singularity. Therefore, $\psi_B$ does not generate flat bands. The other kind of flat bands originates from the {destructive interference of the states on} the dice lattice structure. We can construct these wave functions in the $\alpha=0$ limit, to which the model is continuously connected} \cite{Tarnopolsky2019}, namely, $\Psi_0\propto (ai\bar{\partial}^*\Lambda_1({\bf r}),bi\bar{\partial}^*\Lambda_2({\bf r}),0,0,ai\bar{\partial}\Lambda_1({\bf r}),bi\bar{\partial}\Lambda_2({\bf r}))^T$, where $a$, and $b$  are constants and $\Lambda_{1,2}$ are some arbitrary functions of ${\bf r}$ without singularities. The Chern number of these flat bands is zero. This is also confirmed in the pseudo-Landau-level description discussed in Sec. \ref{seciv}. From the  Landau level point of view, the flat bands corresponding to the zeroth Landau level are topological and the ones from the $n$th $(n>1)$ Landau level are trivial.

\begin{figure}
	\centering
	\subfigure[\label{a}]{\includegraphics[scale=0.1]{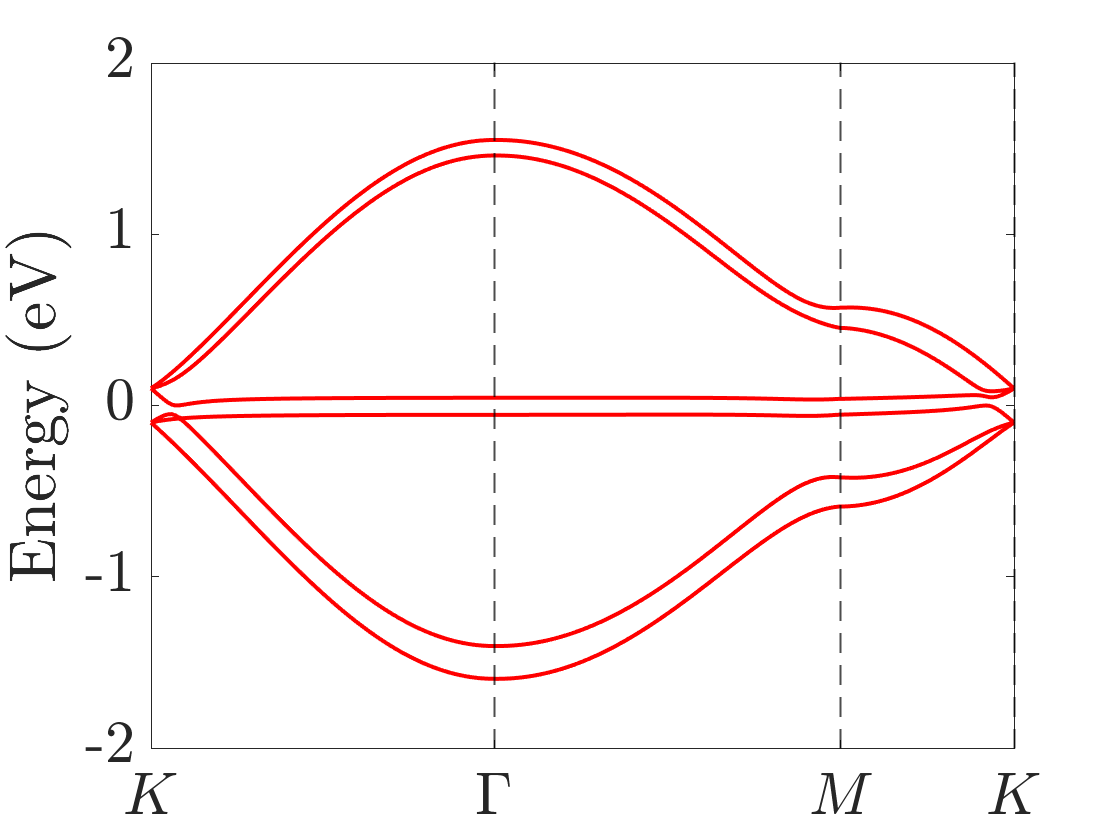}}
	\subfigure[\label{b}]{\includegraphics[scale=0.1]{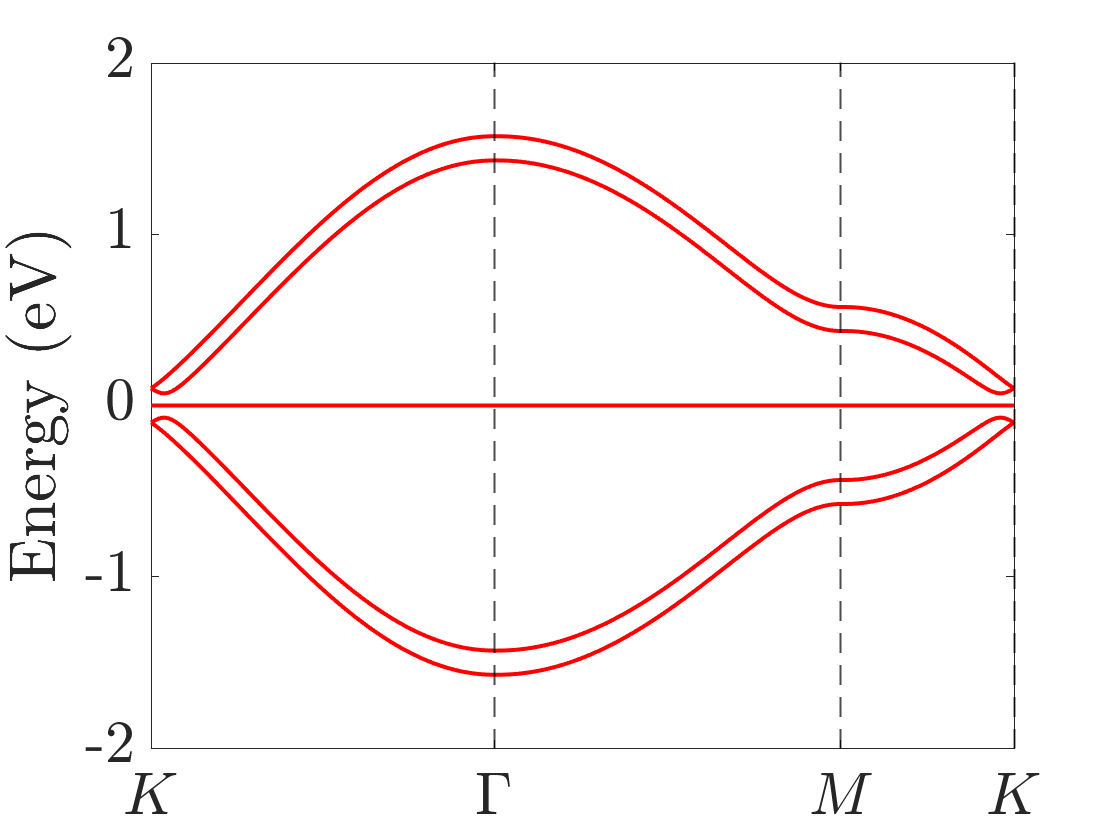}}
	\caption{(color online)
		{
		Band structure of the $A-B$ stacking untwisted bilayer dice lattice model. Here we set the reciprocal path the same as for graphene \cite{catarina2019twisted}; the lattice vector and coordinates of atoms in the unit cell are the same as in Fig. \ref{fig1}(b). The other parameters are $W_{AB'} = W_{BC'} = 0.33$ eV and (a) $W_{AC'}=0.33$ eV (the chiral symmetry is broken) and (b) $W_{AC'}=0$ (the chiral symmetry is preserved)
		}
		\label{fig4}	}
\end{figure}

{The chiral symmetry of the Hamiltonian (\ref{htbd}) can also be confirmed without twisting (see the Appendix). For $A-B$ stacking, finite $W_{AC'}$ will break the particle-hole symmetry [see Fig. \ref{fig4}(a)]. }

\subsection{Comparing the band structures of TBG and the TBD} \label{seciii}

After building the effective model of the TBD,
we now can compare the band structures of TBG and the TBD. {Similar to Bistritzer and MacDonald's model for TBG \cite{Bistritzer-MacDonald2011}, the Hamiltonian for a layer with the twist angle $\theta$ near the Dirac point ${\bf K}_\theta$ can be written as
\begin{equation}
	H_{{\bf K}_\theta}({\bf q}_\theta)=v_fq_\theta\left(
	\begin{array}{ccc}
	0	& e^{-i(\theta_q-\theta)}  & 0 \\
	e^{i(\theta_q-\theta)}	& 0 & e^{-i(\theta_q-\theta)} \\
	0	& e^{i(\theta_q-\theta)} & 0
	\end{array}
	\right), \label{htheta}
\end{equation}
where ${\bf q}_\theta=q_\theta (\cos \theta_q,\sin\theta_q)$ is the momentum measured from the Dirac point ${\bf K}_\theta$ in the moir{\'e} Brillouin zone with the unit vectors $\boldsymbol{b^m}_1 = \frac{4\pi}{\sqrt{3}L_s}(\frac{1}{2}, -\frac{\sqrt{3}}{2})$ and $\boldsymbol{b^m}_2 = \frac{4\pi}{\sqrt{3}L_s}(\frac{1}{2}, \frac{\sqrt{3}}{2})$ where  $L_s = \frac{a}{2\sin (\theta/2)}$ is the size of the {moir{\'e}} unit cell [see Fig. \ref{fig3}(b)]. By using the Bloch bands, we can truncate the TBD model Hamiltonian near the Dirac points in the {moir{\'e}} Brillouin zone. For example, by defining $\boldsymbol{b} = l_1 \boldsymbol{b^m}_1 + l_2 \boldsymbol{b^m}_2$, the simplest case is to truncate to the first  {moir{\'e}} Brillouin zone \cite{Bistritzer-MacDonald2011}, namely $l_1=l_2=1$. The truncated Hamiltonian has the form
\begin{eqnarray}
	&&H_{tr}=\nonumber\\
	&&\left(
	\begin{array}{cccc}
H_{\bf K}^1({\bf q})	& T_{{\bf q}_b}	& T_{{\bf q}_{tr}} & T_{{\bf q}_{tl}} \\
T_{{\bf q}_b}^\dagger	&H_{\bf K}^2({\bf q}+{\bf q}_b)	& 0 & 0 \\
T_{{\bf q}_{tr}}^\dagger	&0	& H_{\bf K}^2({\bf q}+{\bf q}_{tr}) & 0 \\
T_{{\bf q}_{tl}}^\dagger	& 0  & 0 & H_{\bf K}^2({\bf q}+{\bf q}_{tl})
	\end{array}
	\right),\nonumber\\
\end{eqnarray}
where $H_{\bf K}^{1(2)}$ is the kinetic part of (\ref{htheta}) in layer 1(2). The basis of the above Hamiltonian is four three-component spinors with the momentum near the central Dirac point in layer 1 and ${\bf q}_b$, ${\bf q}_{tr}$, and ${\bf q}_{tl}$ in layer 2 [see Fig. \ref{fig3}(b)]. The chiral limit is obtained by replacing the hopping matrices $T_{{\bf q}_b}$, $T_{{\bf q}_{tr}}$, and $T_{{\bf q}_{tl}}$ with $T_{{\bf q}_b}^c$, $T_{{\bf q}_{tr}}^c$, and $T_{{\bf q}_{tl}}^c$, respectively. }  

\begin{figure}
	\subfigure[\label{1a}]{\includegraphics[scale=0.10]{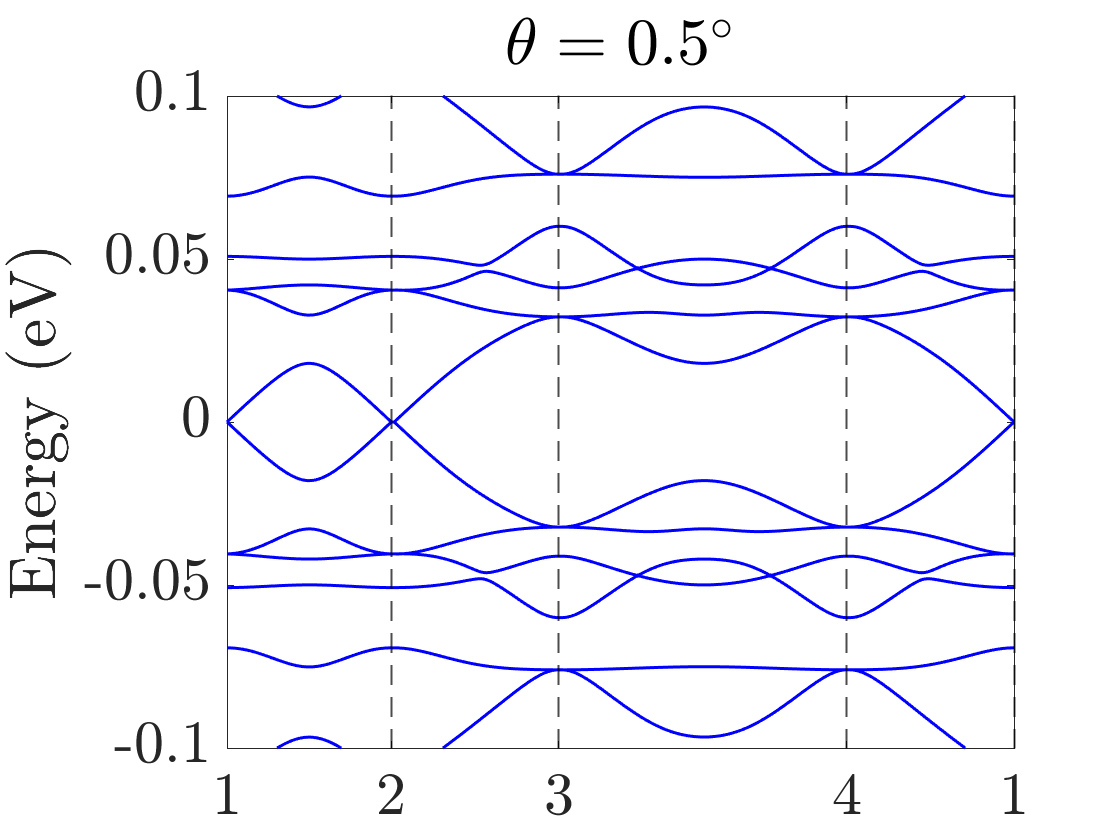}}
	\subfigure[\label{1b}]{\includegraphics[scale=0.10]{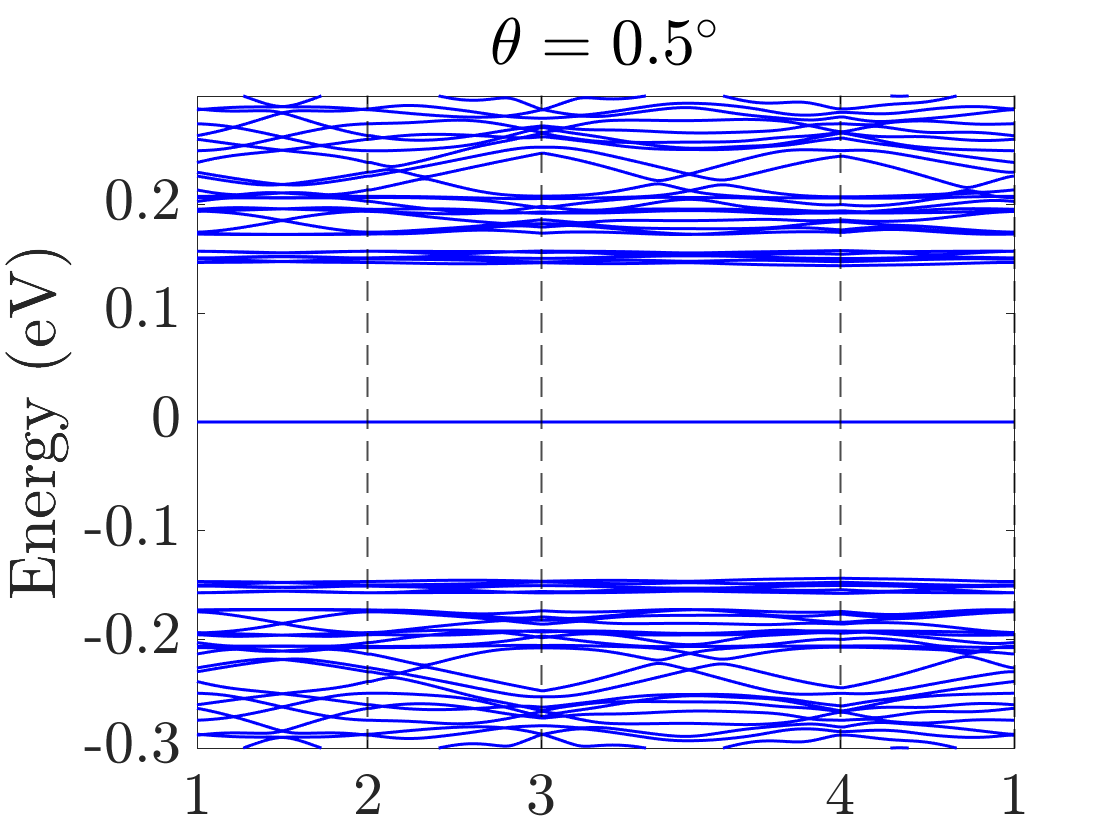} }
	\subfigure[\label{1b}]{\includegraphics[scale=0.10]{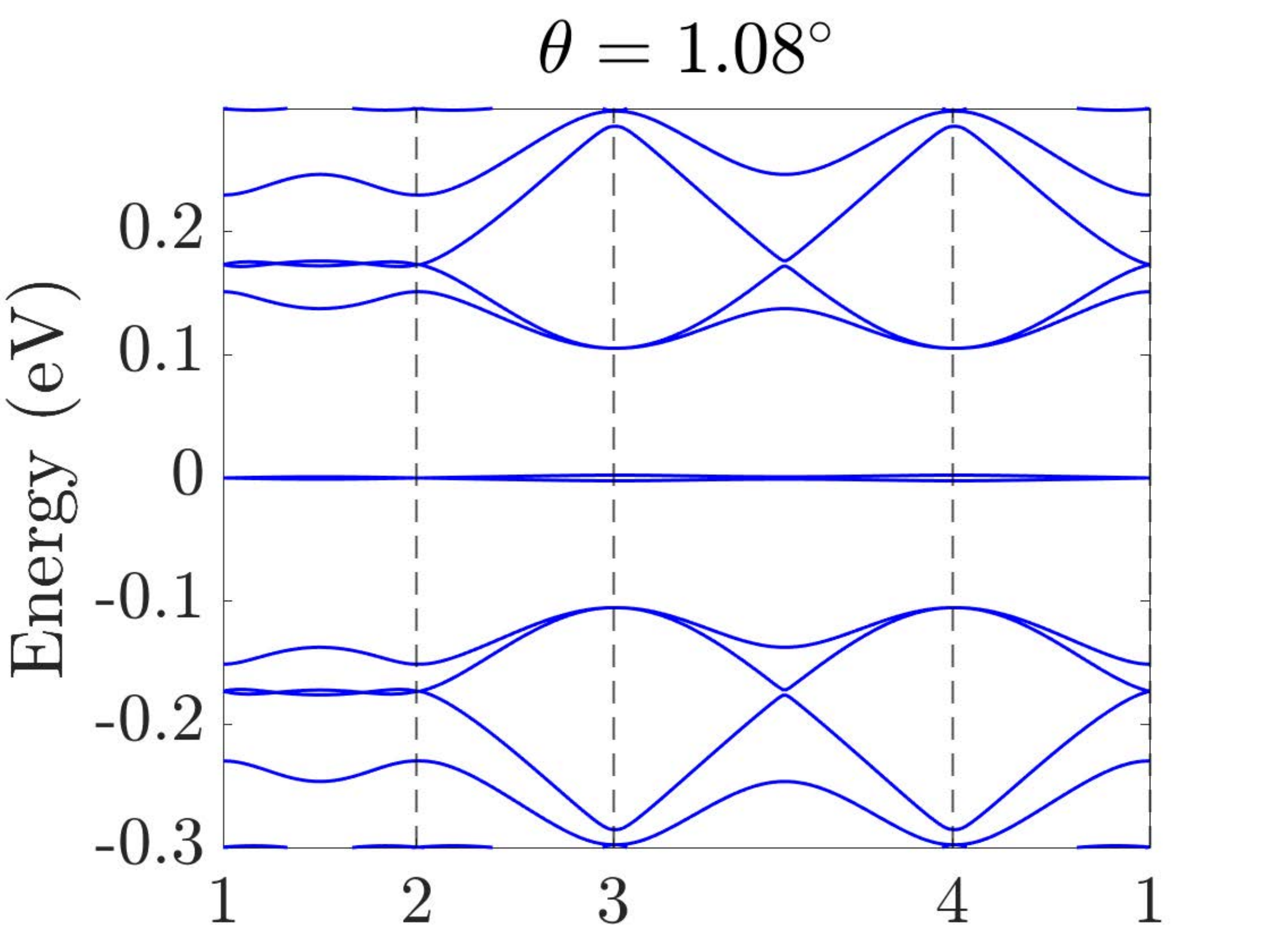} }
	\subfigure[\label{1b}]{\includegraphics[scale=0.10]{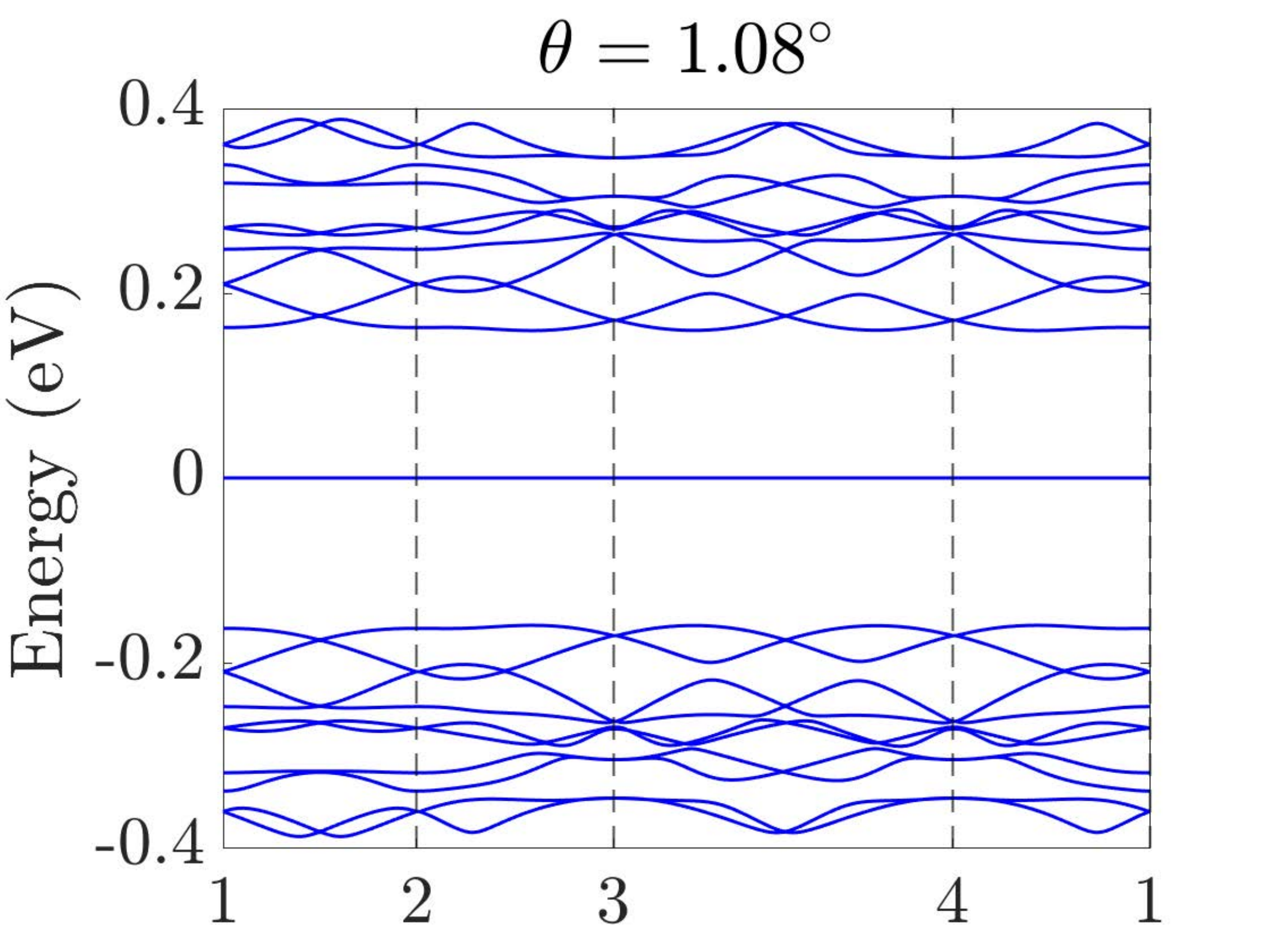} }
	\subfigure[\label{1b}]{\includegraphics[scale=0.10]{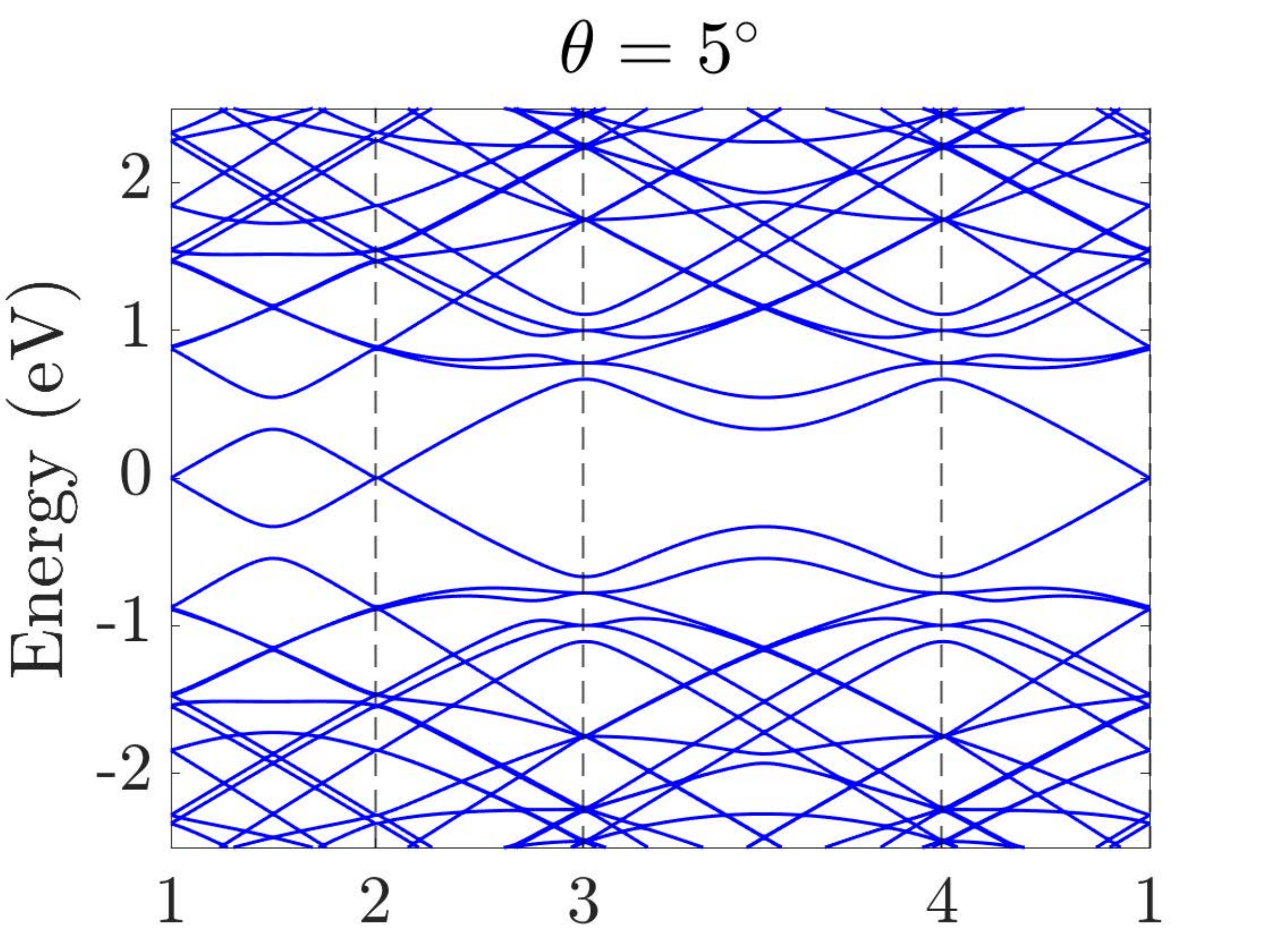} }
	\subfigure[\label{1b}]{\includegraphics[scale=0.10]{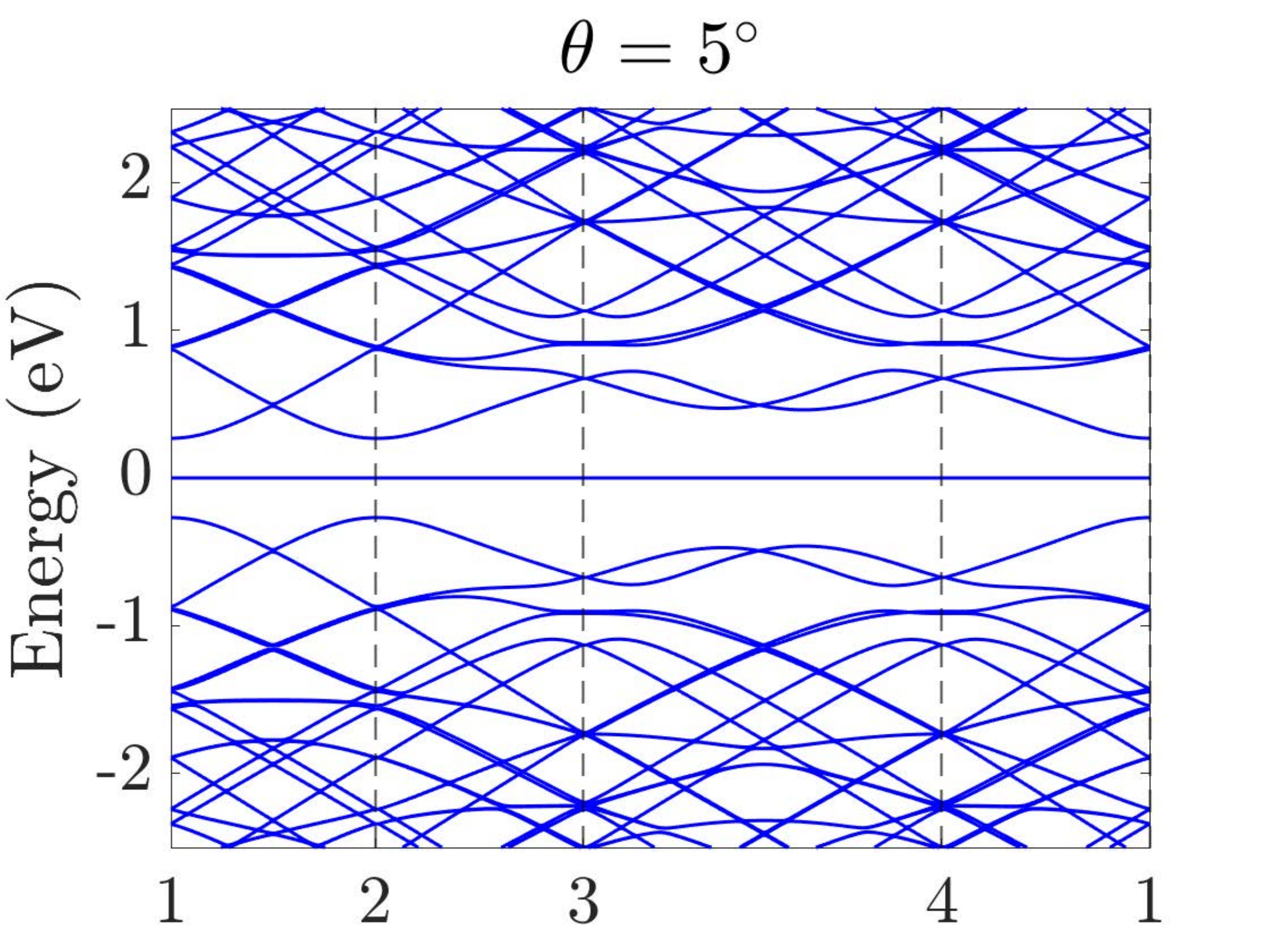} }
	\subfigure[\label{1b}]{\includegraphics[scale=0.075]{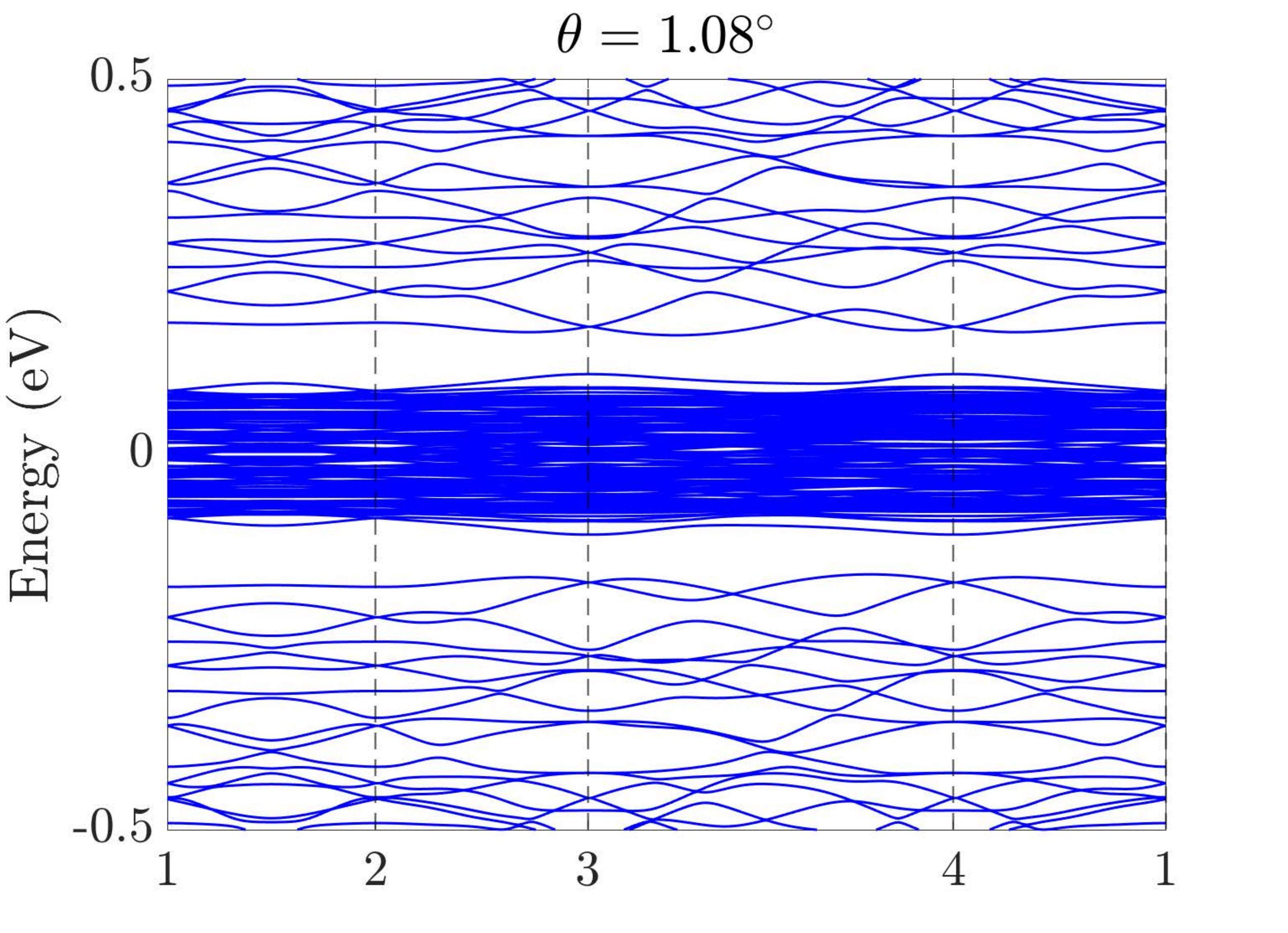} }
	\subfigure[\label{1b}]{\includegraphics[scale=0.10]{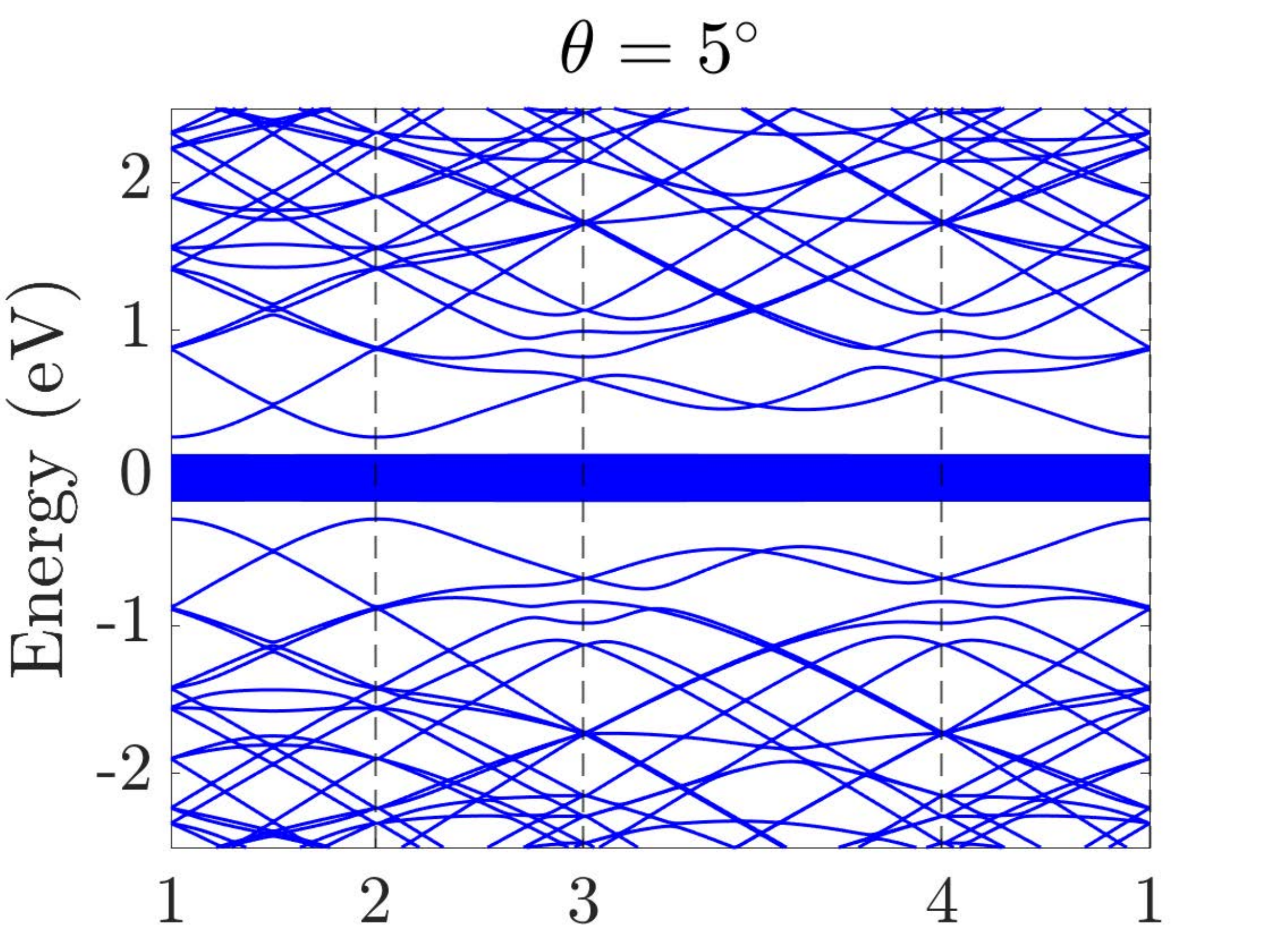} }
	\caption{(color online)
		{Here $W_{AB'}=W=0.11$ eV. The  band structure of TBG in the chiral limit is shown at twist angles (a) $\theta = 0.5^{\degree}$, (c) $\theta =1.08^{\degree}$, and (e) $\theta =5^{\degree}$. The flat band only occurs at the magic angle $\theta =  1.08^{\degree}$. The band structure of the TBD in the chiral limit is shown at twist angles (b) $\theta = 0.5^{\degree}$, (d) $\theta =1.08^{\degree}$, and (f) $\theta= 5^{\degree}$. Totally flat bands with a high level degeneracy
			appear at zero energy at all angles. The band structure of the TBD is shown (g) at magic angle $\theta=1.08^{\degree}$ with $W_{AA'}=0$ eV and $W_{AC'}=0.1$ eV, and (h) at $\theta = 5^{\degree}$, with $W_{AA'}/W_{AB'} = 0.5$. In (g) and (h) the chiral symmetry is broken and the degeneracy is lifted. The momentum labels $1\rightarrow2\rightarrow3\rightarrow4\rightarrow1$ are defined in Fig. \ref{fig3}(b).
		}
		\label{fig2}	}	
\end{figure}

{In the numerical calculation, we choose $W_{AA'} =W_{BB'}=W_{CC'}$.} In TBG, the absolutely flat band at the magic angle $\theta \approx 1.08^{\degree}$ can be obtained in the chiral limit, namely, by choosing $W_{AA'} = 0$ \cite{Tarnopolsky2019}. In the chiral limit of the TBD, highly degenerate flat bands at zero energy also exist. We truncate the Hamiltonian at the order of {$l_1 = l_2 = 3$}, and we choose three {moir{\'e}} angles $\theta = 0.5^{\degree}, 1.08^{\degree}, 5^{\degree}$. The final results of the band structures of TBG and the TBD are summarized in Fig.\ref{fig2}.

Unlike TBG which only has flat bands near magic angles, the TBD has flat bands at all angles, which is a manifestation of the flat band of the single-layer dice lattice model. These TBD flat bands are highly degenerate in the chiral limit, {more than the usual band counting. When chiral symmetry is broken by finite $W_{AA'}$ or $W_{AC'}$,  the original exactly flat bands in the chiral limit now spread from zero energy and become nearly flat; this behavior is similar to that of  Landau levels [see Figs. \ref{fig2}(g) and \ref{Figure5}(b)]. {Numerically, we find that for small perturbations that break the chiral symmetry,} the nearly flat bands are now away from zero energy and the gaps among them and the conduction and valence bands remain {[see Figs. \ref{fig2}(g) and \ref{fig2}(h)]. This, however, could be an artifact of the continuum model due to flat bands and the lack of valley structure, which is the limitation of the continuum model away from the chiral limit. To further verify the validity of the continuum model in the 
flat-band regime, a more controlled calculation, such as a
real-space commensurate one, is needed, which is
beyond the scope of the present work.} Therefore, besides the theoretical analysis in the preceding section, we also numerically confirm that in order to obtain the band structure of exact flat bands at zero energy, namely, the chiral limit, both parameters $W_{AA'}$ and $W_{AC'}$ must be zero.}

	{Before ending this section, we would like to comment on the degeneracy of the flat bands of the TBD. In general, the degeneracy is lower when away from the magic angle or the chiral limit. In numerical calculations, apply different cutoff parameters for different twist angles, because the size of the unit cell dependent on the twist angle $|\bm{A}_{m.u.c.}| = |\bm{a}_1^m \times \bm{a}_2^m | = (3\sqrt{3}d^2)/(8\sin^2(\theta/2))$. The smaller the twist angle is, the more Dirac points will be folded in this cutoff. A consequence is that the number of degenerate flat bands decreases as the twist angle increases. Further, the number of flat bands is nearly one-third the number of sites, showing that most of the flat bands originate from the destructive interference of the states on the dice lattice. Our results are consistent with those in Ref. \cite{arv2023}.}



	
\section{ The $SU(2)$ gauge potential and pseudo-Landau-level structure} \label{seciv}
{The chiral limit is an ideal model to study the flat-band structure, while in reality, the chiral limit is violated by finite $W_{AA'}$ or $W_{AC'}$ arising from different atomic stacking and atomic layer deformation \cite{PhysRevB.90.155451}. Fortunately, {Liu $et$ $al$}. showed that Bistritzer and MacDonald's model for TBG at a small magic angle can be effectively described by pseudo-Landau levels under an $SU(2)$ gauge potential \cite{Liu2019Pseudo}. After a gauge transformation on the Bloch functions and expanding the tunneling potential to the linear order of $r/L_s$, Liu $et$ $al.$ arrived at the pseudo-Landau-level Hamiltonian for TBG, 
 	\begin{equation}
 		H_{TBG}^{p} = -\hbar v_f ( \mathbf{k} - \frac{e}{\hbar}\mathbf{A}\tau_2 )\cdot {\bf \sigma} + 3u_0\tau_{1},
 	\end{equation}
 where $\mathbf{A} = (2\pi u'_0)/(L_s e v_f)(y, -x)$ is the $SU(2)$ gauge potential and the Pauli matrices $\tau_{1,2}$ act on the layer index. In addition, $u'_0$ denotes the hopping parameters $W_{AB'}$ and $u_0$ denotes $W_{AA'}$. The effective magnetic field ${\bf B}=\nabla\times{\bf A}\approx 120$ T for TBG at  $\theta \approx 1.08^{\degree}$ \cite{Liu2019Pseudo,San-Jose2012Non-Abelian,ren2021}. The effective magnetic fields have opposite directions on each layer; therefore, we can define a time-reversal operator $\Theta=i(\tau_2\otimes\sigma_0)\mathcal{K}$, where $\mathcal{K}$ means complex conjugation. In addition,  the {time-reversal symmetry is preserved.} After a similar treatment, we can also derive the pseudo-Landau-level Hamiltonian for the TBD. By replacing ${\bf \sigma}$, which acts on the sublattice index of graphene, by ${\bf S}$, the result is
\begin{gather}
	H_{TBD}^{p} = -\hbar v_f ( \mathbf{k} - \frac{e}{\hbar}\mathbf{A}\tau_2 )\cdot {\bf S} + 3u_0\tau_{1}, \label{eq6}
\end{gather}
where $u'_0$ denotes hopping parameters $W_{AB'}$ and $W_{BC'}$ in the TBD, and $u_0$ denotes $W_{AA'}$. In the following numerical calculations, we set  $u'_0 = 0.01$ eV, and $u_0 = 0.1$ eV.}

To discuss the Landau-level structure, we transform the Hamiltonian (\ref{eq6}) into the basis that diagonalizes $\tau_2$,
\begin{eqnarray}	
	&&H_{TBD}^p(\mathbf{k}) =\nonumber\\
	&&{\scriptsize \begin{pmatrix}
	0& \pi_x + i\pi_y &0 & 3iu_0& 0& 0\\
	\pi_x - i\pi_y& 0& \pi_x + i\pi_y& 0& 3iu_0& 0\\
	0& \pi_x - i\pi_y&0 &0& 0& 3iu_0\\
	-3iu_0& 0& 0&0& \pi'_x + i\pi'_y& 0\\
	0& -3iu_0& 0& \pi'_x - i\pi'_y& 0 &\pi'_x + i\pi'_y\\
	0& 0& -3iu_0& 0& \pi'_x - i\pi'_y& 0
	\end{pmatrix}},\nonumber\\\label{hami}
\end{eqnarray}
where $\pi_x = \hbar v_f k_x - e v_f A_x$, $\pi_y = -\hbar v_f k_y + e v_f A_y$, $\pi'_x = \hbar v_f k_x + e v_f A_x$, and $\pi'_y = -\hbar v_f k_y - ev_f A_y$. Since $u_0$ is small, we can treat it as a perturbation. Then for the unperturbed Hamiltonian $H_0$, we define the Landau-level creation operators of the two layers, $b^\dagger =\sqrt{ \frac{L_s}{8 \pi u'_0 \hbar v_f} }(\pi_x - i\pi_y) $ and $a^\dagger = \sqrt{ \frac{L_s}{8 \pi u'_0 \hbar v_f} }(\pi'_x + i\pi'_y)$, and the nonzero commutators are $[b,b^\dag]=[a,a^\dagger]=1$, 
{
\begin{gather}
	H_0(\mathbf{k}) =\hbar {\omega}\begin{pmatrix}
		0& b &0 & 0& 0& 0\\
		b^\dagger& 0& b& 0& 0& 0\\
		0& b^\dagger&0 &0& 0& 0\\
		0& 0& 0&0& a^\dagger& 0\\
		0& 0& 0& a& 0 &a^\dagger\\
		0& 0& 0& 0& a& 0
	\end{pmatrix},\label{hami0}
\end{gather}
where $\hbar \omega = \sqrt{ \frac{8 \pi u'_0 \hbar v_f}{L_s} }$. The $H_0$ is block-diagonalized for each layer. We denote by $|\Psi^{(1)} \rangle $ and $|\Psi ^{(2)}\rangle$ the eigen wave functions for layers 1 and 2, respectively. The components of $|\Psi^{(1)} \rangle $ and $|\Psi ^{(2)}\rangle$ are made of Landau-level wave functions that satisfy $b\ket{n} = \sqrt{n}\ket{n-1}$, $b^\dagger\ket{n} = \sqrt{n+1}\ket{n+1}$, $a\ket{m} = \sqrt{m}\ket{m-1}$, and $a^\dagger \ket{m} = \sqrt{m + 1}\ket{m+1}$. To be more specific, 
\begin{eqnarray}
	|\Psi^{(1)} \rangle&=&(A_n^1|n\rangle,A_n^2|n+1\rangle,A_n^3|n+2\rangle)^T,\\
|\Psi^{(2)} \rangle&=&(B_m^1|m+2\rangle,B_m^2|m+1\rangle,B_m^3|m\rangle)^T,
\end{eqnarray}
where $A_n^i$ and $B_m^j$ are the normalizing coefficients and when $n(m)<0$, $|n(m)\rangle=0$.}

{For layer 1, there are three energies for $n>1$,
\begin{equation}
	E_{n,\pm}^{(1)} = \pm \hbar {\omega} \sqrt{2n + 1},\quad E_{n,0}^{(1)} = 0.
\end{equation}
In the wave function for $E_{n,0}^{(1)}$, the second component $A_{n,0}^2=0$, which has a structure similar to that in the flat-band wave function of the dice lattice. For $n=1$, there are two eigenenergies
\begin{equation}
	E_{1,\pm}^{(1)}=\pm \hbar \omega,
\end{equation}
which are the first Landau levels, with the wave functions $\Psi_{1,\pm}^{(1)}=\frac{1}{\sqrt{2}}(0,\pm|0\rangle,|1\rangle)^T$. Here $n=0$ is the topological zeroth Landau level with $E_{0,0}^{(1)}=0$ and $|\Psi_{0,0}^{(1)}\rangle=(0,0,|0\rangle)^T$. }

{For layer 2, the derivation for the spectrum and wave function is similar. For $m>1$, $E_{m,\pm}^{(2)} = \pm \hbar {\omega} \sqrt{2m + 1}$ and $E_{m,0}^{(2)} = 0$. The second component of $|\Psi_{m,0}^{(2)}\rangle$, $B_{m,0}^2=0$. For $m=1$, $	E_{1,\pm}^{(2)}=\pm \hbar \omega$ and $|\Psi_{1,\pm}^{(2)}\rangle=\frac{1}{\sqrt{2}}(|1\rangle,\pm|0\rangle,0)^T$. For $m=0$, $E_{0,0}^{(2)}=0$, and $|\Psi_{0,0}^{2}\rangle=(|0\rangle,0,0)^T$.}

\begin{figure*}
	\centering
	\subfigure[\label{a}]{\includegraphics[scale=0.5]{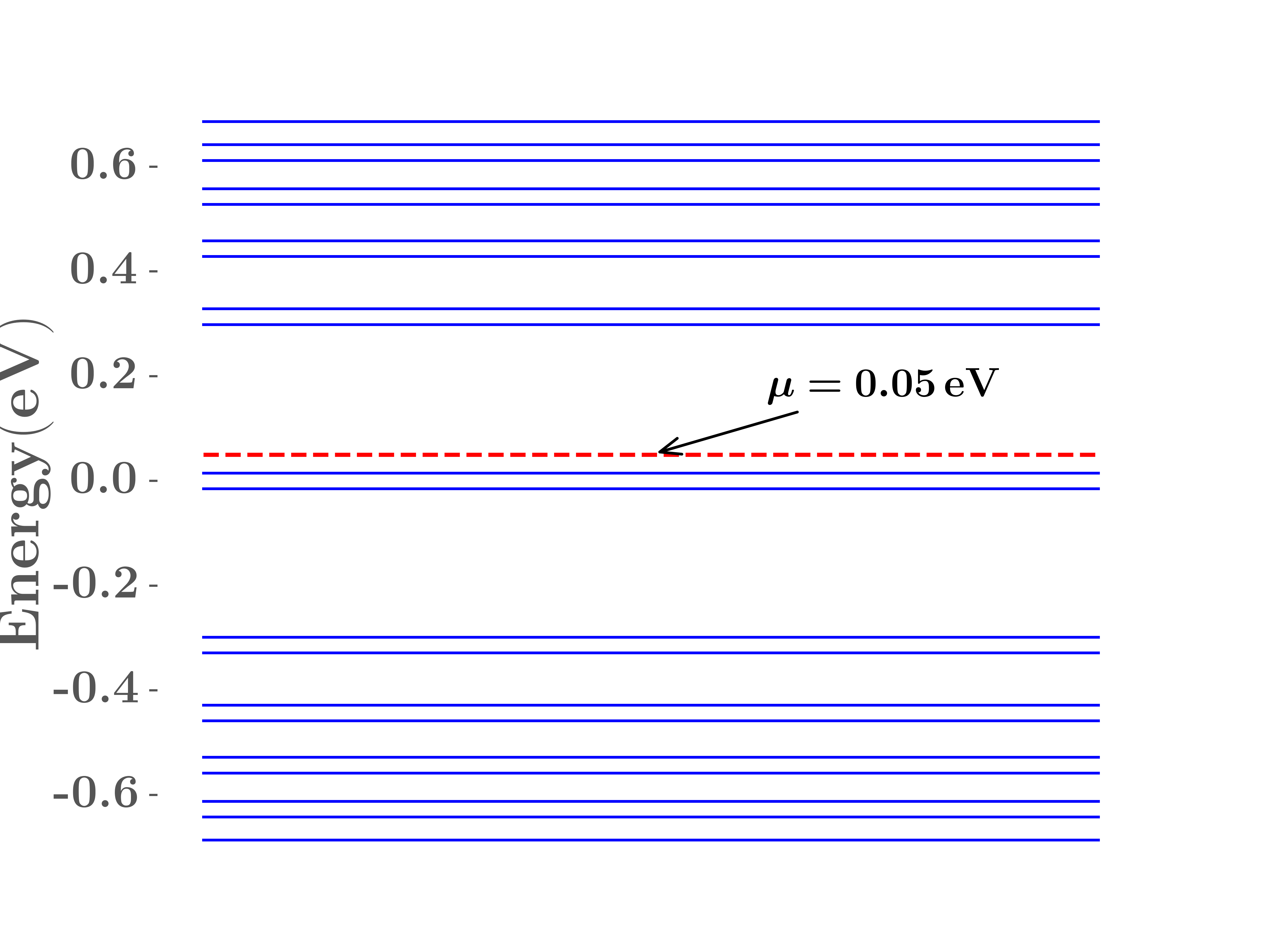}}
	\subfigure[\label{b}]{\includegraphics[scale=0.5]{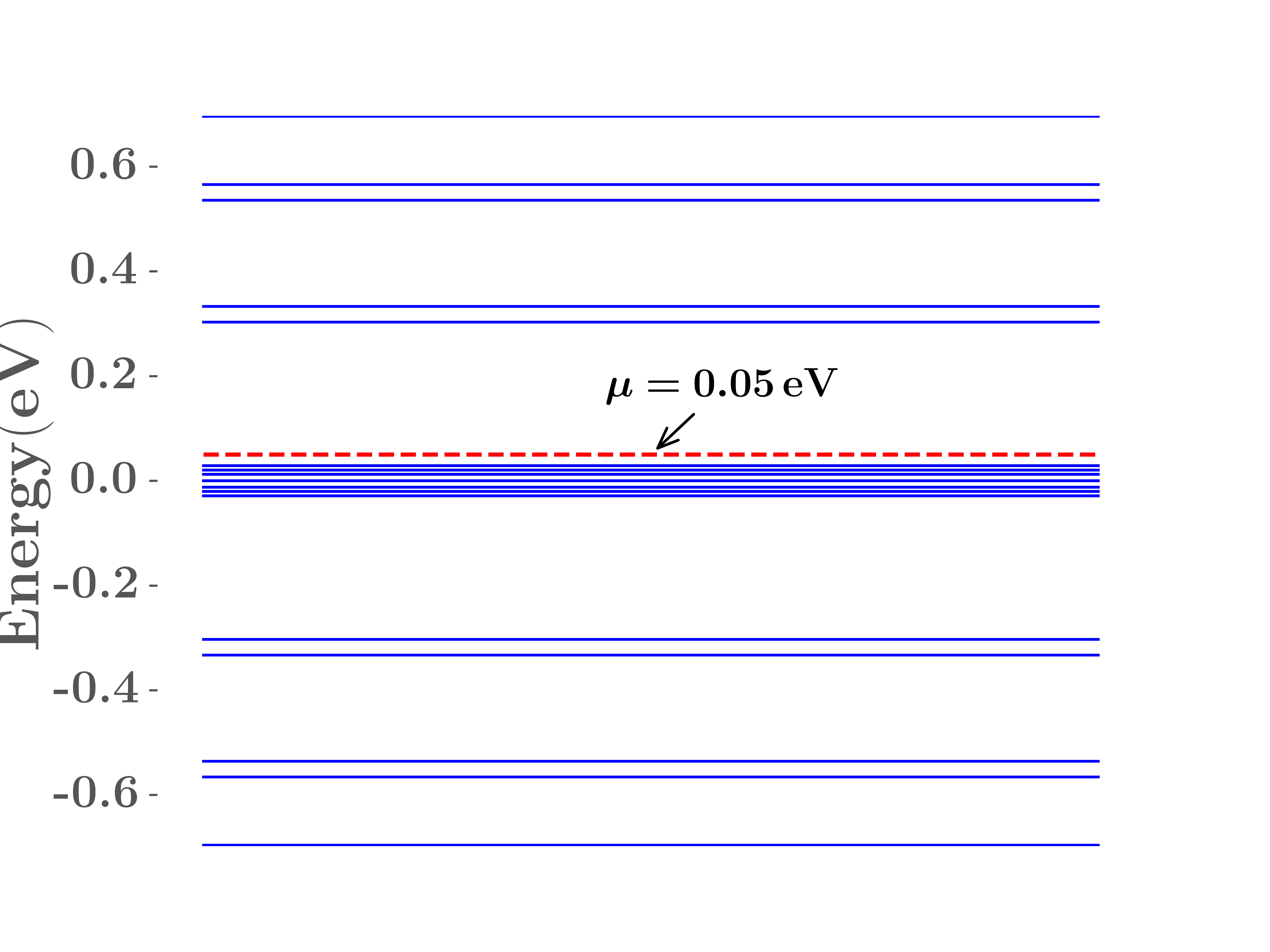}}\\
	\subfigure[\label{c}]{\includegraphics[scale=0.5]{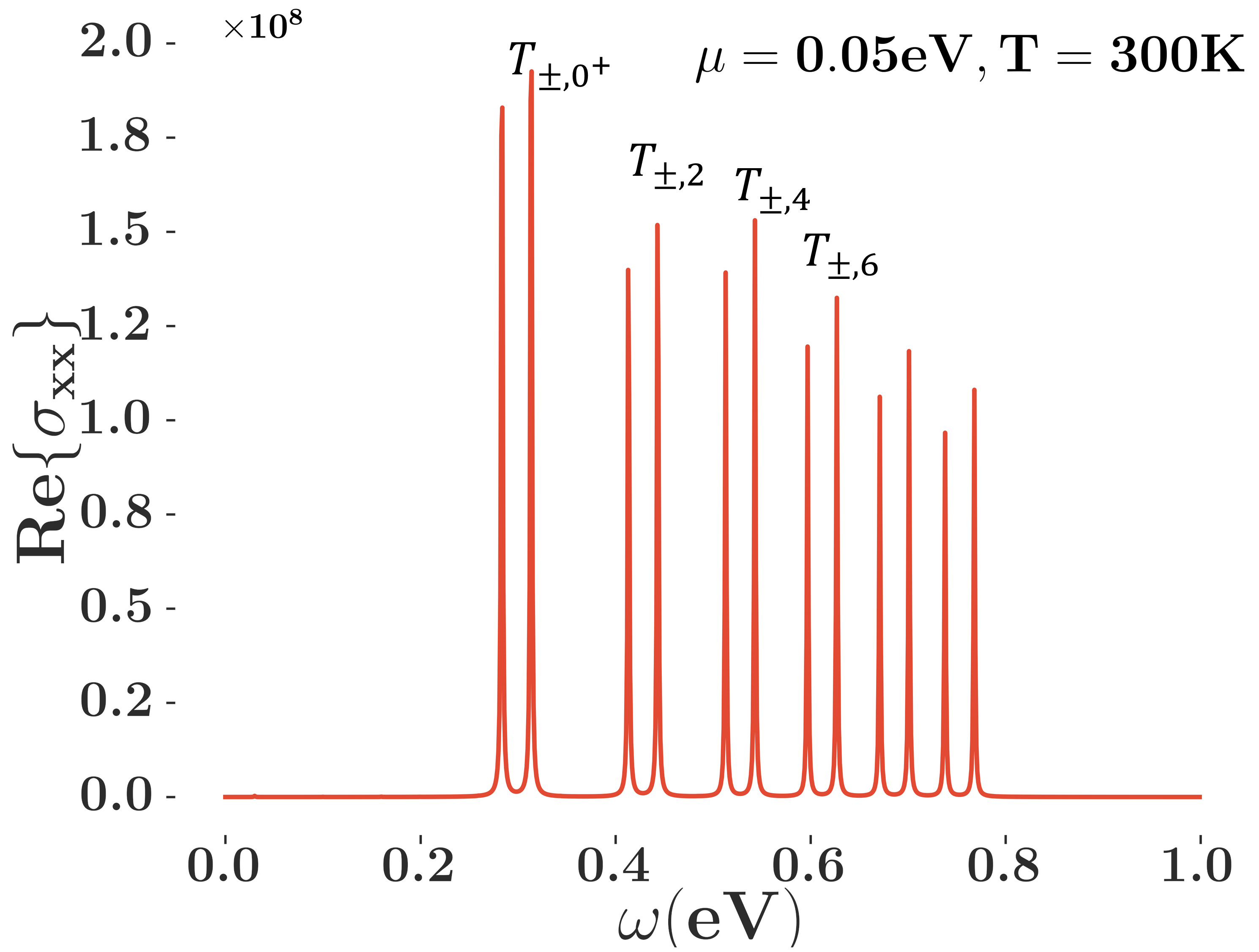}}
	\subfigure[\label{d}]{\includegraphics[scale=0.5]{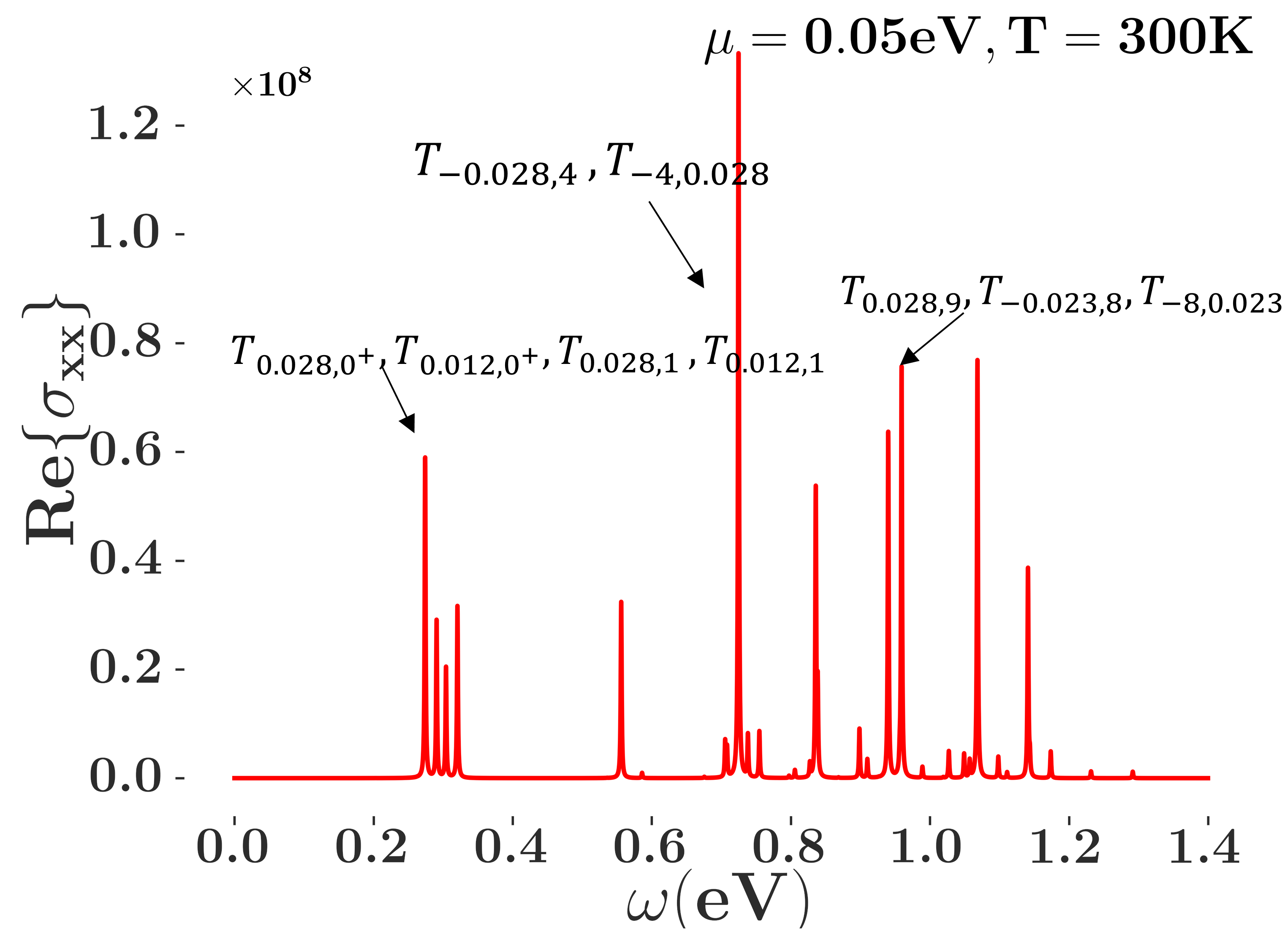}}\\
	\caption{(color online)
		{ 
		Landau-level structure of (a) TBG and (b) the TBD, with $u'_0 = 0.01$ eV, $u_0 = 0.1$ eV, and  $\mu = 0.05 $ eV (red dotted line), and optical conductance $\sigma_{xx}$ of (c) TBG and (d) the TBD. We choose the room temperature $T=300$ K. The symbol $\pm$ in (c) represents the two flat bands near zero energy. For energy bands bigger (or smaller) than the zero energy with a typical Landau-level gap, we denote by $L = 0^{\pm}, \pm1 ,\pm2, ...$. For example, $T_{\pm, 0^+}$ represents the contribution from two zero energy bands to the first Landau level with positive energy denoted by $ L = 0^+ $. For the flat bands near zero energy in (d), we use their energy value to denote, for example, $T_{-0.012,0^+}$, which respects the contribution from the flat band with energy $E = -0.012$ eV to $L = 0^+$.
		}
		\label{Figure5}	}
\end{figure*}
{We treat small $u_0$ as a perturbation, which will mix the states between $|\Psi^{(1)}\rangle$ and $|\Psi^{(2)}\rangle$ and lift the degeneracy at zero energy. This is consistent with the results in Sec. \ref{secii} because $u_0$ corresponds to $W_{AA'}$, which breaks the chiral symmetry. This behavior of small $u_0$ is also confirmed numerically [see Fig. \ref{Figure5}(b)], which is similar to the band structure in Fig. \ref{fig2}. Here we emphasize a difference between TBG and the TBD. Because of the lack of zero-energy states $|\Psi_{n,0}\rangle$ for $n>1$ in the pseudo-Landau levels of TBG, the effect of $u_0$ is of the third order \cite{Liu2019Pseudo}, while in the TBD, it is of the first order. Therefore, the optical conductance for finite $u_0$ will behave differently for TBG and the TBD (discussed below). A similarity between TBG and the TBD is that, after the perturbation of $u_0$, a double degeneracy remains which is related to the $S_3$ symmetry \cite{Liu2019Pseudo}.}

\section{ Optical conductance} \label{secv}

In reality, materials that rotate the polarization plane of linearly polarized light have many applications in various devices, which is usually achieved by magneto-optical effects, the quantum Hall effect, and the Kerr and Faraday rotations by invoking external magnetic fields \cite{oc1,oc2,oc3,oc4,oc7}, which also limits the applications in small-scale devices. Therefore, searching for materials with intrinsic properties that rotate light is urgent for recent applications. The TBG is such a candidate for advanced optical applications due to the tunable twist angles \cite{occ1,occ2}. The optical conductivity of the bilayer dice lattice without twisting was studied before \cite{suk1,suk2}. Here we study the optical conductance of the TBD, which could be another candidate for such applications.

We utilize the Kubo formula in the {pseudo}-Landau-level basis to calculate the optical conductance of the TBD system \cite{PhysRevB.94.125435},
\begin{gather}
	\sigma_{\alpha, \beta} = \frac{ig}{2\pi \hbar l^2_{B}} \sum_{LL_s} 
	\frac{f - f'}{\varepsilon' - \varepsilon } \frac{\braket{\Psi| j_\alpha |\Psi'}\braket{\Psi'|j_\beta|\Psi}}{ \omega - (\varepsilon' - \varepsilon) + i\Gamma  } , \label{sigma}
\end{gather}
where $l_B = \sqrt{(L_s h c v_f )/4\pi u'_0}$ is the magnetic length. Without considering the spin degree of freedom, we can simply set $g = 2$. Here $f$ is the Fermi distribution and $\omega$ is the photon energy. Although we use the {pseudo}-Landau-level description, no {external} magnetic field is applied. {The pseudomagnetic field is induced by the hopping of $W_{AB'}$ and $W_{BC'}$, which can also be induced by strain \cite{guo2022}.}

We take into account the contribution of all Landau levels;
however, the remaining double degeneracy of the band structure will cause divergence {in} conductance. For a doubly degenerate band, we set the divergent part of the Kubo formula  (\ref{sigma}) as
\begin{gather}
 \lim_{\epsilon' - \epsilon \to 0}\frac{f(\epsilon) - f'(\epsilon')}{\epsilon' - \epsilon}  =  \frac{e^{(\epsilon_0 - \mu)/kT}}{(e^{(\epsilon_0 - \mu)/kT} + 1)^2}\frac{1}{kT}.
\end{gather}
From the Hamiltonian {(\ref{hami0})}, the current is obtained as  $j_{\alpha, \beta} = \frac{\delta H_0(\mathbf{k})}{\delta A_{\alpha,\beta}}$.

The numerical results for $\mu=0.05$ eV and $T=300$ K of the conductance of {TBG and} the TBD are shown in {Figs. \ref{Figure5}(c) and \ref{Figure5}(d). Since the chemical potential we choose is close to zero, all significant absorption peaks originate from transitions between the near-zero bands and the positive-energy bands, or between the negative-energy bands and the near-zero bands.} The double-peak structure {of TBG} is caused by 
the splitting of double degeneracy, and the width of the splitting {is proportional to} $u'_0$. 

 {By comparing the result of TBG with that for the TBD, we see that the double-peak structure of TBG does not exist in the TBD. In the TBD, the peak splits into several small peaks. This reflects that there are many states near zero energy. In Fig. \ref{Figure5}(c) the distance between the double peaks is $\Delta\omega\sim 0.03$ eV, which corresponds to 40 $\mu$m, while in Fig. \ref{Figure5}(d) the typical distance between the split peaks is $\Delta\omega\sim 0.015$ eV, and the corresponding wave length is about 80 $\mu$m, which are all in the terahertz range and experimentally detectable. Therefore, this peak-splitting structure provides proof of the experimental prediction showing the existence of the large degeneracy of the flat bands in the TBD. }

Although the Landau-level description for TBG fails when the twist angle is away from the magic ones, the Landau-level structure remains in the TBD [see Figs. \ref{fig2}(g) and \ref{fig2}(h)]. Therefore, in the optical conductance at angles other than the magic ones, the peak splitting structure remains in the TBD, whereas the peaks arising from the transitions that form the flat bands disappear in TBG, which is also a key experimental difference between TBG and the TBD. 

The transitions between these zero-energy levels are forbidden in the TBD, which means there are no peaks at low frequency near $\omega\sim 0$, which is a key difference from the three-dimensional Kane fermion \cite{Luo2019}. The reason for this phenomenon is attributed to the structure of the current and wave function in two dimensions, that is, $\braket{\Psi| j_\alpha |\Psi'}\braket{\Psi'|j_\beta|\Psi} = 0$ between those near-zero bands, and their contribution to the optical conductance being zero, while in three dimensions, the $k_3S_3$ part will have a nontrivial contribution, and the transition between different Landau levels has a non zero $k_3$, which causes the peaks near zero frequency \cite{Luo2019}.

\section{Conclusions} \label{con}

In this paper we constructed a lattice model for the TBD. {In the chiral limit,} it has flat bands at all twisted angles besides the magic ones {and the flat bands are broadened when chiral symmetry is broken}, which could be confirmed by the peak splitting structure of the optical conductance near the magic angles. Away from the magic angles, the peak splitting remains in the TBD, whereas these peaks disappear in TBG due to the nonexistence of the flat bands. The flat bands in the TBD are composed of zero-Chern-number bands by destructive interference of the states on the dice lattice as well as the topological nontrivial bands by the {moir\'e} structure at the magic angles. In this model we have neglected the spin degrees of freedom of electrons. If the spin orbital coupling interaction is added, the bands with zero Chern number may become non-trivial \cite{wang2011}. 
It is possible to realize the TBD in the transition-metal oxide {SrTiO3/SrIrO3/SrTiO3} trilayer heterostructure by growing and twisting in the (111) direction. As a semiconductor, due to the high DOS of the flat bands of the TBD, the TBD may have potential applications in temperature-sensitive and photosensitive manipulations. With interactions, the TBD may also be a good candidate as a fractional Chern insulator.  

{\acknowledgments}
This work was supported by the National Natural Science
Foundation of China through Grant No.~12174067.

\appendix

\section{ Bilayer dice lattice without twisting}\label{app2}

\setcounter{equation}{0}
\renewcommand{\theequation}{A\arabic{equation}}

{In the main text we discussed the effects of finite $W_{AA'}$ and $W_{AC'}$ on the flat-band structure of the TBD system; here we consider their effects  on the band structure in the aligned bilayer case. The Hamiltonian of the $A-B$ stacking bilayer dice lattice reads
\begin{gather}
	H = \begin{pmatrix}
		0& h(\bm{k})& 0& 0& W_{AB'}& W_{AC'}\\
		h^*(\bm{k})& 0& f(\bm{k})& W_{BA'}& 0& W_{BC'}\\
		0& f^*(\bm{k})& 0& W_{CA'}& W_{CB'}& 0\\
		0& W_{AB'}^*& W_{AC'}^*& 0& h(\bm{k})& 0\\
		W_{BA'}^*& 0& W_{BC'}^*& h^*(\bm{k})& 0& f(\bm{k})\\
		 W_{CA'}^*& W_{CB'}^*& 0& 0& f^*(\bm{k})& 0
	\end{pmatrix},
\end{gather}
{where $h(\bm{k}) = -\tau(e^{i \bm{k} \cdot \tau_B} + e^{i \bm{k} \cdot ( \tau_B - a_1 ) } + e^{i \bm{k} \cdot (\tau_B - a_2 ) } )$, $f(k) = -\tau(e^{i \bm{k} \cdot (\tau_B - \tau_C	) } + e^{i \bm{k} \cdot (\tau_B - a_1 - \tau_C) } + e^{i \bm{k} \cdot (\tau_B - a_2 - \tau_C) } )$, and $\tau =\sqrt{2} v_f \hbar/(3d)$.} The basis of the aligned bilayer dice lattices is $\Psi^\dagger = (c^\dagger_{1,A}, c^\dagger_{1,B}, c^\dagger_{1,C}, c^\dagger_{2, A'}, c^\dagger_{2,B'}, c^\dagger_{2, C'})$. The positions of atoms $B$ and $C$ in one unit cell are $\tau_B = d(0, 1)$ and $\tau_C = d(0, 2)$, and $d$ is a lattice constant. Here we have set $W_{AA'}=W_{BB'}=W_{CC'}=0$. The band structures with $W_{AC'} \neq 0$ and $W_{AC'} = 0$ were plotted in Fig. \ref{fig4}, where the chiral symmetry is broken for finite  $W_{AC'}$.}

\bibliographystyle{apsrev}

\bibliography{savedrecs(1)}

\end{document}